\documentclass[5p,times]{elsarticle}
\usepackage{amssymb}
\usepackage{subfig}
\usepackage{afterpage}
\usepackage[table]{xcolor}

\usepackage{amsmath}
\usepackage{multirow}
\usepackage[graphicx]{realboxes}
\journal{FCGS}%Knowledge-based Systems}

\begin{document}

\begin{frontmatter}



\title{Hack Me If You Can: %Using 
%Attention-based 
Aggregating AutoEncoders for Countering Persistent Access Threats Within Highly Imbalanced Data.}


\author[inst1]{Sidahmed Benabderrahmane *}

\affiliation[inst1]{organization={Division of Science},%Department and Organization
            addressline={New York University}, 
            city={Abu Dhabi}, 
            country={UAE}}

\author[inst1]{Ngoc  Hoang}
\author[inst3]{Petko  Valtchev}
\author[inst2]{James  Cheney}
\author[inst1]{Talal  Rahwan}
\affiliation[inst2]{organization={The University of Edinburgh},%Department and Organization
            addressline={School of Informatics, 10 Crichton Street}, 
            city={Edinburgh}, 
            country={UK}}
\affiliation[inst3]{organization={UQAM},%Department and Organization
            addressline={University of Quebec in Montreal, H3C 3P8}, 
            city={Montreal (Quebec) }, 
            country={Canada.},\\
           mail={ ~~~~*: sidahmed.benabderrahmane@gmail.com}}
%^{*}: sidahmed.benabderrahmane@gmail.com
\begin{abstract}
Advanced Persistent Threats (APTs) are sophisticated, targeted cyberattacks designed to gain unauthorized access to systems and remain undetected for extended periods. To evade detection, APT cyberattacks deceive defense layers with breaches and exploits, thereby complicating exposure by traditional anomaly detection-based security methods. The challenge of detecting APTs with machine learning is compounded by the rarity of relevant datasets and the significant imbalance in the data, which makes the detection process highly burdensome.
We present AE-APT, a deep learning-based tool for APT detection that features a family of AutoEncoder methods ranging from a basic one to a Transformer-based one. We evaluated our tool on a suite of provenance trace databases produced by the DARPA Transparent Computing program, where APT-like attacks constitute as little as 0.004\% of the data. The datasets span multiple operating systems, including Android, Linux, BSD, and Windows, and cover two attack scenarios. The outcomes showed that AE-APT has significantly higher detection rates compared to its competitors, indicating superior performance in detecting and ranking anomalies.\\
%\newline
%\newline
\textbf{Data and code}: https://github.com/ae-apt/AE-APT
\end{abstract}


\end{frontmatter}


\section{Introduction}
\subsection{Overview:}
Over the past few years, there has been a significant increase in the popularity of connected devices, enhanced by a large capacity of data accessibility, connectivity, and versatility. As a result, governments, companies, non-commercial organizations or private users are increasingly relying on information technologies (IT) to carry out particularly sensitive activities using large volumes of data, such as e-commerce, asset sharing, information system management, or confidential data processing. This puts the IT platforms directly in the line of fire of cybercriminals, who are continuously scrutinizing exploits for a wide range of information-gathering, data stealing and damaging~\cite{sood2012}. Consequently, cybersecurity has become a major concern which in turn motivated a booming activity aimed at protecting networks, confidential data and vital organization information to prevent them from getting into the wrong hands through cybermenaces~\cite{sujeetha2019cyber,humayun2020cyber}.
An advanced persistent threat (APT) is a typical case of cybermenace:  
An APT is by definition a sophisticated cyberattack campaign in which an intruder establishes a long-term, illicit access on a network in order to steal highly sensitive data~\cite{chen2014study,alshamrani2019survey, JIA2023110781}. The targets of such attacks, which are very carefully chosen, typically include large enterprises or governmental networks, bank and finance entities, or defense agencies~\cite{cole2012advanced, BREWER20145,app13063409, SARHAN2021107524}. These types of attacks are carefully planned and designed to infiltrate the organizations by evading the existing security measures and ``flying under the radar''~\cite{ghafir2014advanced}. The consequences of such intrusions are vast and may include: intellectual property theft, compromise of sensitive information, sabotage of critical organizational data, or total website takeover. For instance, in the case of intellectual property theft, the perpetrators are likely being able to siphon thousands of gigabytes worth of sensitive proprietary information from technology and manufacturing companies around the world. These  usually state-sponsored actors design what is known as a house-of-cards style infection chain to exfiltrate massive troves of highly sensitive data, such as trade secrets or technology patents~\cite{halbert2016intellectual,shackelford2016protecting}. To compromise sensitive information, the attackers could be tempted to get access and damage power distribution grids, telecommunications utilities,  administrative infrastructure systems, employees or users' private data, media archives, electoral and other political targets. In particular, organized crime groups may sponsor APTs in order to gain information they can use to carry out criminal acts for financial gain~\cite{ussath2016advanced}. In the case of critical organizational infrastructures sabotage, database deletion is usually performed under the umbrella of hacktivism. Indeed, hacktivists engage in a disruptive or damaging activity on behalf of a cause, be it political, social or religious in nature. These individuals or groups often see themselves as virtual vigilantes, working to expose fraud, wrongdoing or corporate greed, draw attention to human rights violations, protest censorship or highlight other social injustices~\cite{tankard2011advanced}.
Finally, during  website destruction, the hackers use smart techniques to overpass the encryption methods that are used to hide the web data content, by analyzing patterns in the network traffic data, and then utilize these patterns to shut down the websites~\cite{chen2021few}.

Performing an APT penetration requires more resources than a classical web application attack, since a higher degree of skills is needed in the customization and sophistication of the attacks. APT hackers are usually part of a team of experienced cybercriminals with substantial financial backing. Some APT attacks are government-funded, and used in cyberwarfare. %\\
APTs differ from traditional web application threats, in that:
(i) They are significantly more complex. (ii) They are not \textit{hit-and-run} attacks: Once a network is infiltrated, the intruder remains within in order to extract as much information as possible.
 (iii) They are executed against a specific mark and indiscriminately launched against a large pool of targets.
(iv) They often aim to infiltrate an entire network, as opposed to one specific part thereof.

Depending on its exact method, an APT attack may leave multiple traces and signs. Beside phishing-email campaigns, these could include: 
    (a) unusual activities on user accounts, such as an increase in high-level logins late at night, 
    (b) widespread presence of backdoor Trojans, 
    (c) unexpected or unusual data bundles, which may indicate that data has been amassed in preparation for exfiltration, 
    or (d) unexpected information flows, such as anomalies in outbound data or a sudden, uncharacteristic increase in database operations involving massive quantities of data.
    
Protecting information systems against APT attacks is yet another challenging goal. There have been many attempts in designing cybersecurity and intelligence solutions to assist organizations in better protecting against APTs, in particular using Intrusion Detection Systems (IDS). Some of them suggest the utilization of network firewalls\footnote{Security devices designed to monitor Internet traffic and filter out the undesirable part thereof.}.
Other organizations implement manual threat-hunting policies which boil down to security experts tracking potential threats within the networks on a 24/7 basis. Partnering with a cybersecurity service provider is also a possible solution: Should the attack happen; security firms may bring assistance in response to the threats.

    While the above-mentioned IDS solutions could yield some substantial benefits in terms of anti-APT defense, most of them suffer on a potentially invalidating drawback. It is rooted in a crucial aspect that should not be neglected while planning the security and protection activities: It consists in the potentially forbidding costs of all the necessary the temporal, human and hardware resources. Indeed, in cybersecurity the most important factor is speed: For an effective defense, the protection activities and counter-measures should be executed faster than the attacking ones are performed by the adversaries. The breakout time is measured by how long it takes for an intruder to start infiltrating a network after gaining access to it. It becomes then obvious that a fast and automated process, mimicking human intelligence, is essential.

\subsection{Contribution:}
In this paper we present a new deep learning-based approach that boils down to anomaly detection (AD). Our model learns to discriminate specific APT-like patterns from trace databases. To deal with highly imbalanced datasets (malicious attacks account for a tiny proportion of system activities), it utilizes an AutoEncoder neural net which learns a low-dimensional representation of normal activity data. Any data that substantially differs from that representation in the decoding space is then deemed  anomalous and flagged for closer inspection.
Our method has been implemented and tested using real provenance data of several operating systems, and compared to many existing anomaly detection approaches, where it has shown better detection rates compared to those methods.
The major contributions of this paper are: (i) a deep learning-based pipeline for the detection of APTs, using various AutoEncoder architectures; (ii) design of a baseline AutoEncoder and five variations thereof (adversarial, recurrent, long short-term memory variant, gated recurrent units variant, attention-based); (iii) design an ensemble learning mechanism on top of AutoEncoder-based classifiers;
(iv) 
experimental evaluation of the proposed models using large APT databases;
(v)
comparison of these models to several baseline approaches.

The novelty of our work lies in the development of a platform-agnostic deep learning-based anomaly detection method specifically tailored to detect Advanced Persistent Threats (APTs) within highly imbalanced datasets. By employing an ensemble of advanced AutoEncoder architectures, including baseline, adversarial, recurrent, and attention-based models, our approach achieves significantly higher detection rates compared to traditional methods. Additionally, the incorporation of the self-attention mechanism in transformers further enhances the robustness and accuracy of our detection pipeline. This research contributes to the literature by providing a comprehensive evaluation of these models using real provenance data from multiple operating systems, demonstrating superior performance in identifying APT-like patterns. Cyber-security professionals, IT departments, and organizations across various sectors—including government, finance, and critical infrastructure—will greatly benefit from this work as it offers a powerful tool to enhance their defenses against sophisticated and persistent cyber threats.

The remaining of this paper is organized as follow: section 2 illustrates 
some anomaly detection techniques that have been presented in the literature to counter-attack the sophisticated APTs. Section 3 presents the background and architecture of the proposed deep learning model. Section 4 summarizes the used datasets, the evaluation metrics and the experimentation results. Section 5 concludes the paper with the main outcomes and finally Section 6 gives some future perspectives.
\section{Anomaly Detection for APT Tracking: Related Work}
\begin{figure}
\vspace{-6 em}
    \centering
   \includegraphics[width=1\linewidth]{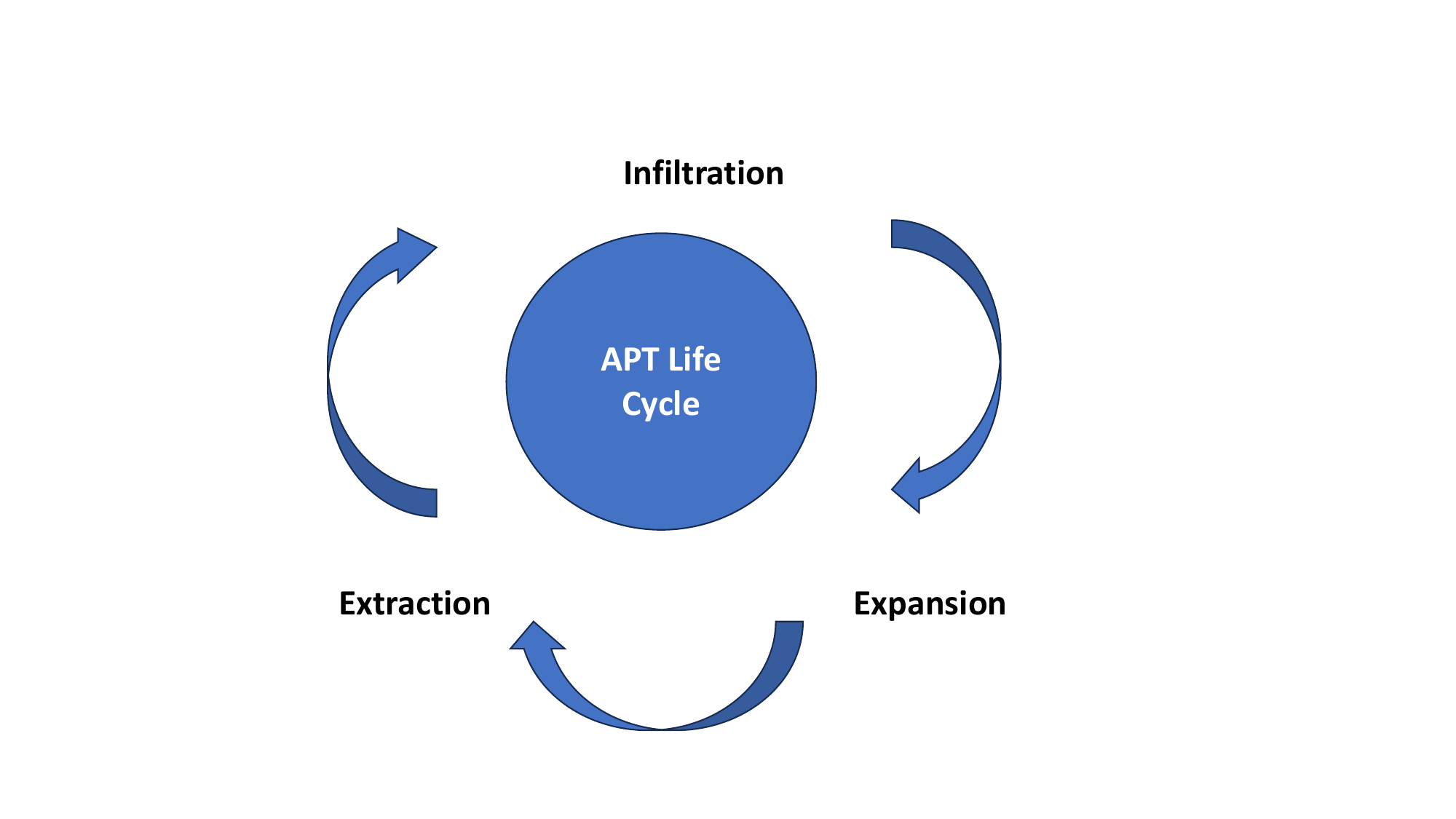}
    \caption{The three main steps of the APT life-cycle: From network infiltration and the expansion of the attacker’s presence to the extraction of the data.}
    \label{fig:life-cycle}
\end{figure}

APT attacks usually utilize customized zero-day malware and exploits such as SQL injection, Remote File Inclusion (RFI), Cross-Site Scripting (XSS), Trojans and back-doors for targeting a specific organization~\cite{kshirsagar2023towards,bhimireddy2023web,BREWER20145}. 
The life-cycle of a successful APT attack can be broken down into three stages: (1) network infiltration, (2) the expansion of the attacker’s presence and (3) the extraction of amassed data—all without being detected~\cite{FRIEDBERG201535,XU2022848}. 
Figure~\ref{fig:life-cycle} illustrates these main three stages of the APT life-cycle.\\
During the past two decades, there have been a notable examples of APT and incursions targeting both governmental organizations and private institutions. Among the most famous APT attacks we can cite: \textit{Goblin Panda APT27}, \textit{Fancy Bear APT28}, \textit{Cozy Bear APT29}, \textit{Ocean Buffalo APT32}, \textit{Helix Kitten APT34}, \textit{Wicked Panda APT41} \textit{APT19}, \textit{Stuxnet}, \textit{Deep panda}, \textit{Epic Turla}, \textit{Oldsmar}, or more recently \textit{Pegasus} \cite{Stuxnet11,marczak2018hide}. This last cyber warfare tool has been capable of reading text messages, tracking calls, collecting passwords, location
tracking, accessing the target device’s microphone and camera,
through a zero-click exploit~\cite{saad2020attribution}. 

Enhancing current intrusion detection systems to enable their detection of APT attacks is a major challenge in both anomaly detection and cybersecurity~\cite{Sakthivelu23}. 
An IDS
is a software intended to identify and classify malicious activities in an information system. The concept, introduced by Denning~\cite{denning1987intrusion}, covers a variety of tools whose functionalities may include, beside mere intrusion detection, further steps ranging from triggering warnings to actively preventing the attackers from causing further harm~\cite{VIEGAS2017200}.
For the last few years, machine learning methods have been extensively studied as tools for network intrusion detection in IDS performing traffic monitoring.
This became possible due to the proliferation of generic Anomaly Detection (AD) approaches as well as to  active research on applications thereof in a variety of contexts: 
sequence data analysis, fraud detection, intrusion detection, medical and health anomaly detection, image processing, and textual data anomaly detection~\cite{aggarwal2017introduction}. 
Within an AD approach, to identify anomalous events/behavior, a model uses a learning database to detect patterns characterizing what is to be considered normal and/or abnormal. As per~\cite{chandola2009anomaly}, three types of anomalies arise in cybersecurity: point-based, contextual-based, and collective-based ones. In point-based AD methods, every single event that deviates from normal behavior can be considered as a point anomaly. In contrast, within contextual AD, the abnormal status of an event can only be established with respect to its context of occurrence. This must be reflected in problem formulation. An anomalous event can then be identified given by the mismatch of its behavioral features within the corresponding context (which might turn out not to be anomalous in a different context). For instance, running  operations such as reading the password files, or accessing the root folder from a \textit{root} account may be normal while performing them while enjoying restricted rights is clearly a risky endeavor. Finally, in collective AD a series of events, or patterns of such events, may be seen as anomalous, whereas the same events, when taken individually, could be assessed as normal. 

Also, in~\cite{chandola2009anomaly}, the authors distinguish six types of AD algorithms that could be applied for detecting the aforementioned categories, namely: classification-based (e.g. deep learning, neural networks, Bayesian networks, support vector machines and implication rules), nearest neighbor-based, clustering-based, statistic-based, information theory-based, and spectral-based AD approaches~\cite{FRIEDBERG201535,chandola2009anomaly}. Classification AD techniques attempt to classify events as being either normal or abnormal using a labeled training data. Nearest neighbor AD methods use the assumption that normal events occur in a close and dense neighborhood of the feature space, thus a metric has to be found to serve as similarity measure for events. Clustering AD methods generate subgroups of highly similar events, with the assumption that anomalies either belong to tiny clusters or to no cluster at all. Information theory AD approaches assume that anomalous events can be identified due to point-wise irregularities in the information content of the learning dataset. They try to generate a model of normality about the information in the data in order to detect local inconsistencies. Finally, spectral AD methods perform a dimensionality reduction by assuming that there exists a subspace where normal data and anomalies are easily separable. All above methods can be explored and tested separately, or combined in a hybrid model to increase the detection rates, depending on the input data and/or the nature of the network/system events. 

Authors of~\cite{Sakthivelu23} relied on random forests, Adaboost, LogitBoost, and logistic regression classifiers for intrusion detection. The \textit{TRAbID} datasets\footnote{https://secplab.ppgia.pucpr.br/?q=trabid} was used to train their ensemble classification models. However, their tool has been designed to analyze a single class of netflow activities, i.e. Remote Desktop protocol (RDP) event logs, as used by only one operating system (OS), Windows, hence its ability to extract abnormal feature sets is limited.
In~\cite{FRIEDBERG201535}, the authors proposed a framework for collecting and analyzing traffic data to identify key phases of intrusion activities corresponding to data exfiltrations. The framework learns a model of normal system behavior over time in the form of a set of rules over log-events. In performance mode, it reports all actions that differ from the model. The framework validation study relied on three synthetic datasets generated following the approach in~\cite{6890935}.
Since their method had no prior knowledge about similarities that might exist between event classes, this  led to the generation of similar log events in redundant hypotheses that overloaded the model. 

In~\cite{MARTINLIRAS2021102202}, authors report on the test of several classification techniques such as SVM, random forest, kNN and regression models for the detection of APT signatures. The dataset they used, made of Windows log files, contains the PE32 (executable file) format, object code, DLLs and others. It uses a legacy 32-bit representation as opposed to the actual 64-bit one in recent Windows versions.  
In~\cite{lamprakis2017unsupervised}, an approach for the analysis of HTTP traffic logs was proposed. The authors deem that protocol the key means used by attackers to establish a command-and-control channel to infected hosts in a network. As a first step, the proposed method extracts the request graph from Web traffic logs. Next, it enhances that graph by click detection and link completion. Finally, the method filters out the remaining unrelated requests that are not blocklisted as anomalous events.

Recently, a deep learning model for APT detection was proposed in~\cite{abdullayeva2021advanced}, which boils down to adding extra layers to an AutoEncoder neural network. The main difference with our own study is in the validation scope: The approach was only tested against a single modestly-sized dataset representing malware for Windows OS. Moreover, the dataset \footnote{https://marcoramilli.com/2016/12/16/malware-training-sets-a-machine-learning-dataset-for-everyone/}
is published in a personal blog and lacks vital information about its provenance (how was it created and then curated?). Correspondingly, the dataset has not been widely used in practical studies\footnote{As of today, in three publications on Google-Scholar}.

Similarly,~\cite{app12136816} proposed a deep learning-based approach that revolves around an AutoEncoder architecture. Again, it was experimentally tested on a single dataset that was produced by combining two separate sources, Contagio\footnote{https://contagiodump.blogspot.com/} and CICIDS2017\footnote{https://www.unb.ca/cic/datasets/ids-2017.html}. As a well-known security benchmark, CICIDS2017 encompasses both benign and network intrusion data that resemble real-world traffic whereas Contagio, reportedly, contains data on effective APT attacks. Only the latter are targeted by the model whereas the attacks from CICIDS2017 are regarded as unrelated to an APT. Unfortunately, as the Contagio dataset is unavailable --as of today-- on the indicated URL (again, a personal blog unrelated to any well-established institution), none of the paper's claims could be verified. Therefore, we see the comparison with this approach as pointless.

Yet another AD approach based on AutoEncoder architectures was proposed in~\cite{9496635}. The presented validation study uses the aforementioned CICIDS2017 and another widely popular benchmark, KDDCUP99 \footnote{http://kdd.ics.uci.edu/databases/kddcup99/kddcup99.html} (neither includes APT attacks). Indeed, more than 50\% of the intrusion detection papers use the DARPA/KDD datasets due to their availability~\cite{AHMED201619net}. However, both datasets are criticized by some for the generation procedure they relied on~\cite{mchugh2000testing}.
Moreover, the analysis in~\cite{mahoney2003analysis} found evidence of simulation artifacts that could result in over-estimations of AD performances. The KDD datasets were developed using a Solaris-based operating system to collect a wide range of data due to its easy deployment. However, significant differences exist with modern OS which barely resemble Solaris. In this age of Ubuntu, Windows and Mac OS, Solaris has almost no market share. The traffic collector used in KDD datasets, TCPdump, is very likely to become overloaded and drop packets from a heavy traffic load. More dramatically, there is some confusion about attack distributions in these datasets. According to an attack analysis~\cite{ALJAWARNEH2018152}, Probe\footnote{Probing: This type of attack collects information of target system prior to initiating an
actual attack.} is not an attack unless the number of iterations exceeds a specific threshold. Furthermore, label inconsistencies have been reported. A description of the KDD datasets states that there are 24 training and 14 test attacks. However, it is reported in~\cite{shafi2013evaluation} that the training data contain 22 attacks and 17 in the test data. This inconsistency has a significant impact on the class distribution of attacks. In this scenario, it is important to create intrusion detection datasets in modern-day computing to address the issues of DARPA/KDD. The next sections discus one such contemporary dataset for network traffic analysis and APT attacks.

Notwithstanding the impressive number of dedicated frameworks, preventing all APT attacks is virtually impossible, experts warn~\cite{auty2015anatomy}. Rather, they advocate continuous system monitoring in order to detect APT at the earliest and thus keep the ensuing damage to a minimum. Yet APT mimicking normal user activity might prove hard to detected with traditional means (e.g. antivirus software, signature or system-policy-based techniques). Non robust machine learning methods relying on system/event logs or audit trails would typically also fail as they only analyze short event/system call sequences: This not only prevents them from properly modeling and capturing long-term behavior patterns but also makes them susceptible to evasion techniques~\cite{streamspot,han2018tapp,unicorn,BerradaCBMMTW20}. That is the case with the above- mentioned classification approaches, where the majority of the available databases are either outdated or of small size, or related to a single OS (typically Windows), hence the reported findings are hardly generalizable.
In fact, method assessment using insufficiently large data and short event sequences is among the major drawbacks of the existing proposals. 
An even more critical one lays in the reliance on private databases which drastically curbs reproducibility.

A recent study presented in~\cite{BerradaCBMMTW20} and~\cite{Benabderrahmane21}, explored the improvement of  APT detection brought by whole-system provenance-tracking and provenance trace mining. The authors argue that richer contextual information of provenance should help identify causal relationships between system activities~\cite{awad2016data,jenkinson2017applying}. These, in turn, enable the detection of attack patterns (e.g. data exfiltration) that usually go unnoticed with the usual perimeter defense-based or policy-driven tools.   
In~\cite{Benabderrahmane21}, an extensive comparative study of AD-based approaches for APT detection is also presented. It exploits a recent, heterogeneous and high-quality provenance databases provided by DARPA's \verb|ADAPT| \\ project~\cite{BerradaCBMMTW20} (see below). Six methods have been benchmarked with that dataset.
Two of them, Valid Frequent Association Rule Mining-based anomaly detection (VF-ARM) and Valid Rare Association Rule Mining (VR-ARM) are contributed by that paper. VF-ARM relies on the extraction of the frequent itemsets from the database, then flags the objects as anomalies when they violate the frequent association rules. On the other hand, the VR-ARM algorithm detects the objects that satisfy at least one rare association rule in the database.
Attribute Value Frequency (AVF)~\cite{koufakou_2007} is a non-parametric approach to outlier detection that has been proven to be scalable, fast, and accurate for categorical data. AVF is a straightforward approach relying on the intuition that outliers in a given categorical dataset are those where the attribute values are infrequent, where the frequency of a certain attribute value is measured across the whole dataset. An AVF score is computed for each data point, where a lower AVF score indicates more infrequent, or unusual, behavior. Frequent Pattern Outlier Factor (FPOF)~\cite{he_2005} focuses on discovering frequent patterns, or itemsets, in the given data and outliers are assumed to be those whose itemsets contain less frequent patterns. As such, discovered frequent patterns are considered the common features of the data, and any data instances that deviate from these patterns are considered anomalous. Outlier Degree (OD)~\cite{narita_2008} also relies on itemset frequency coupled with high-confidence association rules to give scoring to the data points reflecting their potential to be outliers. Similar to FPOF, OD considers data points that do not exhibit frequently-seen behavior to be more anomalous than others. One-Class Classification by Compression (OC3)~\cite{smets2011} works by first identifying frequent itemsets and then finding some subset of itemsets that can provide a good compression of the data. The key idea behind OC3 is that normal items include frequently observed behavior and thus will compress well. Anomalies, on the contrary, will be harder to compress.

In this paper, we present a new deep learning framework (AE-APT) that implements a family of AutoEncoders for APT classification and detection. The training and evaluation of the model have been performed using provenance traces belonging to the aforementioned \verb|ADAPT| provenance project~\cite{BerradaCBMMTW20}, that contain
APT-like attacks on several different host operating systems (Windows, BSD, Linux, and Android) produced as part of the DARPA \verb|Transparent Computing TC| program\footnote{https://www.darpa.mil/program/transparent-computing}. To the best of our knowledge the \verb|ADAPT| project\footnote{https://gitlab.com/adaptdata} is the unique publicly available source that contains databases with long log-event/patterns and, most importantly, traces of different source OS. 
Moreover, overall, the attacks constitute as little as 0.004\% of the data, making proper APT identification a very challenging, since highly imbalanced, classification problem.
In this context, our AE-APT model has been experimentally compared with the AD methods from~\cite{Benabderrahmane21} on the \verb|Transparent|\\ \verb|Computing TC| databases. 
Noteworthily, we chose to ignore too small-sized, obsolete or insufficiently heterogeneous datasets as well as the APT detection methods, however recent, that have been validated \textit{exclusively} on such sources. 

\section{AE-APT: A new AutoEncoder based APT Detection Model}
\subsection{Methodology}

Deep learning has been a great driving force in a variety of computing fields, introducing advanced techniques that utilize neural network architectures to yield state-of-the-art results. In particular, deep learning has been used  in the realm of AD to learn important feature representations, driven by anomaly scores. A number of the methods proposed in this context have outperformed the more conventional methods on a range of real-life challenges and applications~\cite{pang_2020}. Deep learning-based AD methods split into three conceptual categories:
\begin{itemize}
    \item Deep learning for \textit{feature extraction}: Deep learning components of methods in this category work as dimensionality reduction techniques. They aim at extracting low-dimensional feature-based representations of the initial, higher-dimensional data so that AD can be conducted in latter stages. Thus, dimensionality reduction remain a step disjoint from anomaly scores assignment.
    \item Learning \textit{feature representations of normality}: Methods in this category seek to combine feature extraction and anomaly scoring into a single process, rather than keeping them disjoint. They learn some form of expressive representations of normality from the initial, non-transformed data.
    \item End-to-end \textit{anomaly score learning}: The methods in this category combine techniques from the above two categories to create end-to-end systems capable of learning the eventual anomaly scores.
\end{itemize}

\begin{figure}[!htb]
%\vspace{-2 em}
     \centering
     \includegraphics[width=\linewidth]{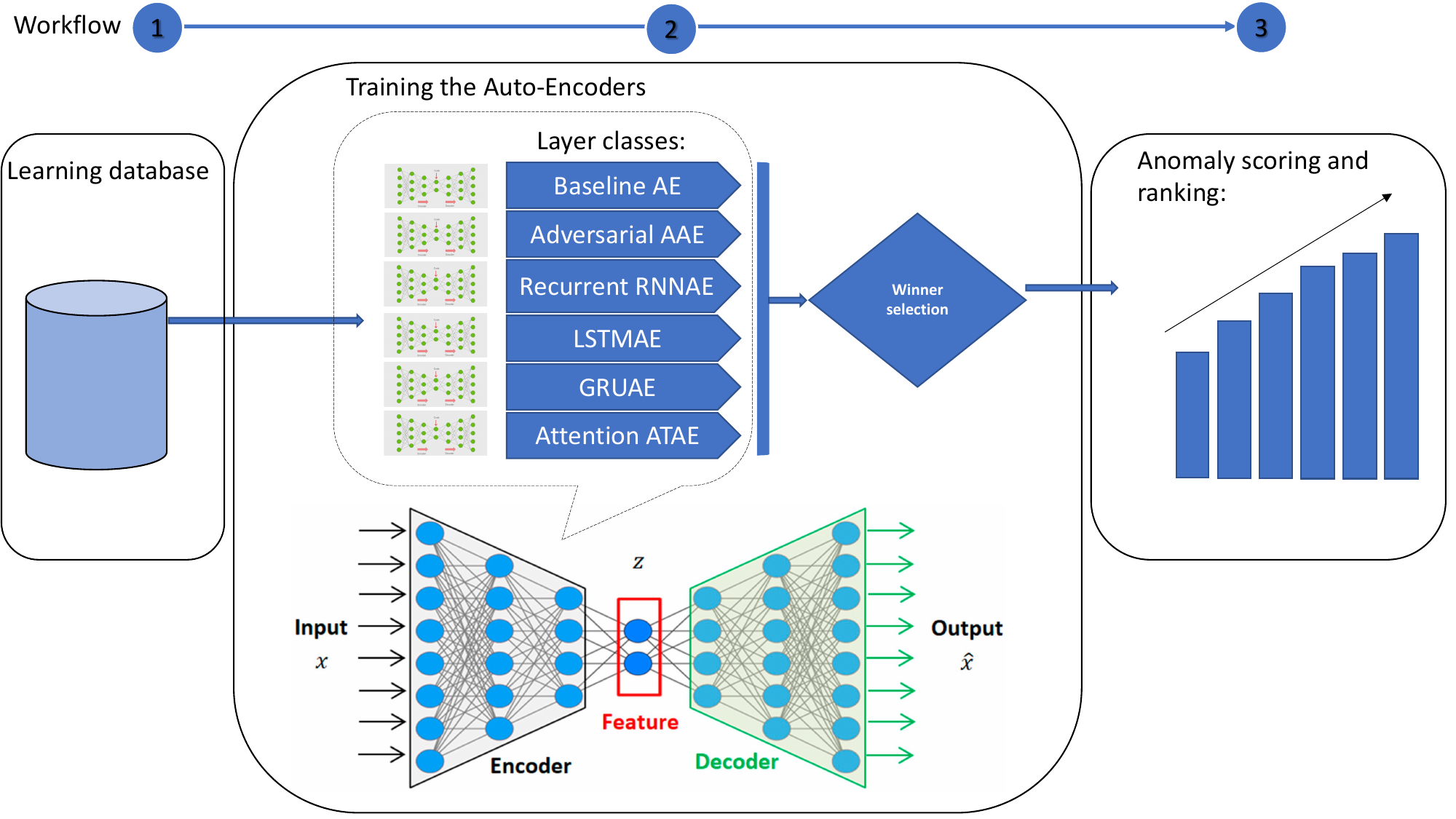}
     \caption{Global architecture of the proposed pipeline AE-APT. Six neural models are trained in parallel, and a winner is selected yielding to the best ranking. }\label{Fig:AEAPT}
\hfill \hfill \hspace{5mm}
\end{figure}

This paper proposes an ensemble classification-based approach for APT detection, called  AE-APT, whose overall architecture is shown in Figure~\ref{Fig:AEAPT}. Its core is made of six neural networks, i.e. a Baseline AutoEncoder (AE) plus five variants thereof: Adversarial (AAE), Recurrent Neural Network (RNNAE), Long Short-Term Memory (LSTMAE), Gated Recurrent Units (GRUAE), and Attention-based (ATAE). Their anomaly scores are combined using a majority aggregation technique. 

Here the target neural models are trained exclusively on normal data, hence they learn meaningful compressed representation of the underlying regularities. The trained models can then be used to --as faithfully as possible-- reconstruct the normal data and thereby discriminate anomalies. We believe this to be highly suitable in the context of APTs where malicious attacks account for a very small proportion of all system activities.

\subsection{Model Selection}
In AE-APT, we propose to use a set of AutoEncoders, as an ensemble, for anomaly detection. Each AutoEncoder tries to reconstruct the data from the input. Then the AutoEncoder with the highest reconstruction error is selected, provided that these miss-reconstructed data points are anomalous.
In our case, it involves training a family of 
six variants of the AutoEncoder architectures.
The final ranking and anomaly scoring is then produced with this winner AutoEncoder.
In what follows, we present them in detail.

\subsection{Baseline AutoEncoder (AE)}\label{autoencoder}
\subsubsection{Background}
The baseline architecture of an AutoEncoder (AE) falls into the second category of the above list: The corresponding AD techniques aim at learning some low-dimensional space for feature representations in which normal data can be reconstructed with minimal reconstruction errors. In doing so, the learned representations are driven to capture essential regularities of the normal data. Anomalous data, in contrast, would be harder to reconstruct from a compressed space and thus will result in greater errors~\cite{pang_2020}.

\subsubsection{Architecture}
\begin{figure}[!htb]

     \centering
     \includegraphics[width=\linewidth]{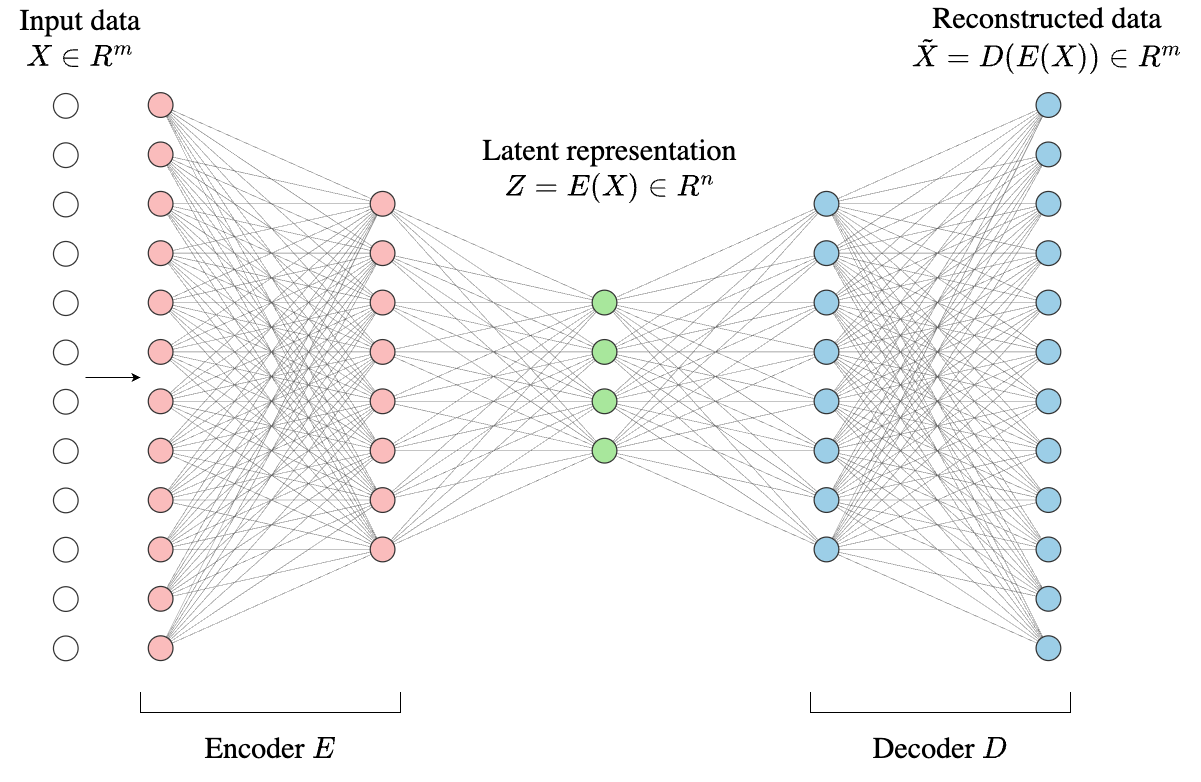}
     \caption{General architecture of the baseline AutoEncoder model (AE).}\label{Fig:AE}
\hfill \hfill \hspace{5mm} 
     \centering
      
     \includegraphics[width=\linewidth]{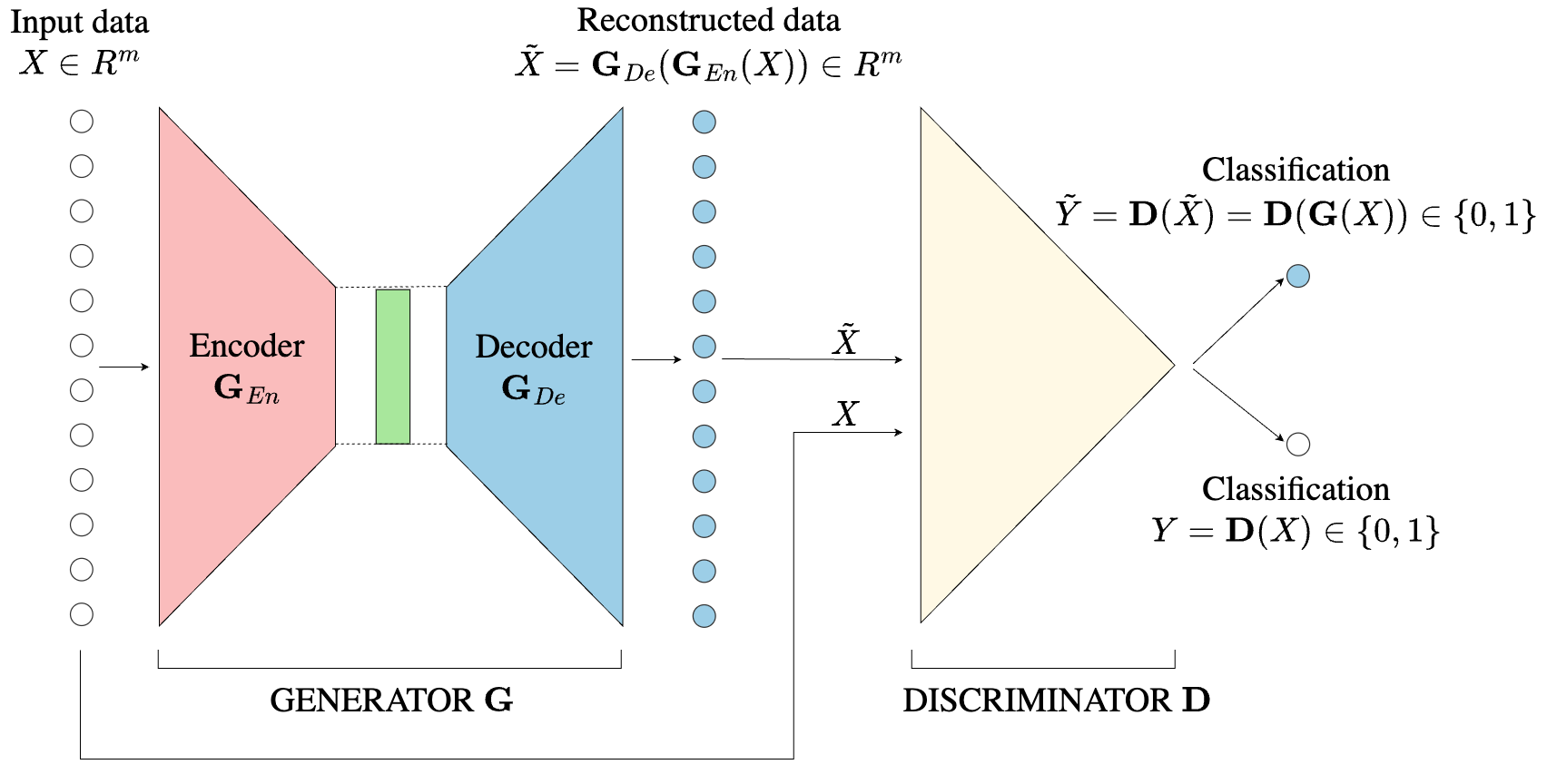}
     \caption{General architecture of the Adversarial AutoEncoder model (AAE).}\label{Fig:AAE}
 
\end{figure}

As per Figure~\ref{Fig:AE}, the baseline AE architecture is made of an encoding network, $\mathbf E$, and a decoding one, $\mathbf D$. The encoder maps the original data of dimension $m$ onto a feature space of dimension $n$ where $n << m$, whereas the decoder reconstructs the original data from its encoding in the feature space.

\subsubsection{Loss function}
A common loss function drives both networks in updating their respective parameters: It reflects the reconstruction error, i.e., the difference between the original data $x$ and the reconstructed data $\tilde{x}$. Formally,
$$L_{AE} = |x - \tilde{x}| = |x - D(E(x))|$$

\subsubsection{Anomaly score}
Through training, the two networks learn to capture the most important compressed features of the normal data points that can be used to reconstruct the data as accurately as possible. As a result, whenever presented with data points with abnormal patterns and features, the trained networks will have more difficulties in correctly reconstructing that data. Consequently, such attempts will generate larger reconstruction errors. Therefore, the reconstruction error of a data point $x$ after being passed through the AutoEncoder is also assigned as its anomaly score:
$$A(x) = |x - \tilde{x}| = |x - D(E(x))|$$

\subsection{Adversarial AutoEncoder (AAE)}
\subsubsection{Background}
The Adversarial AutoEncoders model (AAE) is inspired by the Generative Adversarial Network (GAN) architecture~\cite{goodfellow2014}, which is made of two subnetworks: a generator $\mathbf{G}$ and a discriminator $\mathbf{D}$ with adversarial objectives. The task of $\mathbf{G}$ is to generate data points that most closely match some input distribution, while $\mathbf{D}$ has to distinguish between these points and genuine input data, i.e. drawn from that (true) distribution. At training time, $\mathbf{G}$ is bound to improving its production of realistic data points that make it ever harder for $\mathbf{D}$ to distinguish them from real ones. In other terms, it aims at maximizing the discriminator’s error in labeling data (genuine vs generated).
The rationale behind AAEs is their higher effectiveness in learning meaningful compact representations of input data compared to baseline AutoEncoders. Indeed, through adversarial training, AAEs are led to learn a latent space that tightly fits the underlying structure and variation in the training data. Thus, beside the reconstruction tasks which is core in AutoEncoders, AAEs also excel in generating data samples that closely resemble the training set. 
\textit{This generative capability is facilitated by the adversarial training component, which encourages the model to generate realistic samples that cannot be distinguished from real data by an adversarial discriminator.}

\subsubsection{Architecture}

The proposed AAE model (see Figure~\ref{Fig:AAE}) expands the use of AutoEncoders for AD: In an adversarial training setup similar to that of GANs, $\mathbf{G}$ is a classical AutoEncoder. In this context, the real versus faked data dichotomy of a GAN morphs into the original-to-reconstructed data opposition. This is precisely the task of $\mathbf{D}$, a deep neural network to be trained in distinguishing the original distribution (normal data) from the one reconstructed by $\mathbf{G}$. The idea is that when set into this adversarial training loop, the AutoEncoder is guided towards generating data that closely resemble the original ones. This is achieved through the additional goal of tricking $\mathbf{D}$ into confounding original and reconstructed data. Thus, the training process is driven by a two-fold loss function forcing it to minimize the reconstruction error while maximizing the discriminating one.

The structure and function of $\mathbf{G}$ remain the same as described in section~\ref{autoencoder}: It maps the input data $X$ into $\mathbb{R}^m$ onto a compressed representation ($\textbf{G}_{En}(X)) \in \mathbb{R}^n (n << m)$ and then map this representation back onto the $\mathbb{R}^m$ space to produce the reconstructed data $\tilde{X} = \mathbf{G}_{De}(\mathbf{G}_{En}(X)) \in \mathbb{R}^m$. The discriminator $\mathbf{D}$ is structured as a deep feedforward neural network whose input dimension is also $m$. It is trained on two sets of input data: the set of original data and the set of reconstructed data. For each data point $x$, $\mathbf{D}$ outputs a binary number $\mathbf{D}(x)$ which is 1 whenever $x$ is deemed part of the original input $X$ and 0 otherwise ($x$ assumed in $\tilde{X}$). 

\subsubsection{Loss functions}
Given the aforementioned goal for $\mathbf{D}$ to correctly separate input into original and reconstructed, its loss function is stated as follows:

\begin{equation*}
    \begin{aligned}
        L_{D} &= L_{original} + L_{reconstructed}\\
        &= |\overrightarrow{1} - Y| + |\overrightarrow{0} - \tilde{Y}|\\
        &=|\overrightarrow{1} - \textbf{D}(X)| + |\overrightarrow{0} - \textbf{D}(\tilde{X})|\\
        &=|\overrightarrow{1} - \textbf{D}(X)| + |\overrightarrow{0} - \textbf{D}(\textbf{G}(X))|
    \end{aligned}
\end{equation*}

In parallel, $\mathbf{G}$ is trained both to reconstruct the input data faithfully and fool the discriminator. Therefore, its loss function is formulated as follows:
\begin{equation*}
    \begin{aligned}
        L_G &= L_{reconstruction} - \lambda \times L_D\\
        & = |X - \tilde{X}| - \lambda \times L_D\\
        &= |X - \textbf{G}(X)| - \lambda \times (|\overrightarrow{1} - \textbf{D}(X)| + |\overrightarrow{0} - \textbf{D}(\textbf{G}(X))|)
    \end{aligned}
\end{equation*}
where $\lambda$ is a parameter moderating the impact of the adversarial training. For simplicity, we have set $\lambda$=0.5. 

\subsubsection{Anomaly score}
The anomaly score of a data point $x$ is set to the reconstruction error at the output of the generator:
$$A(x) = |x - \tilde{x}| = |x - \textbf{G}(x)| = |x - \textbf{G}_{De}(\textbf{G}_{En}(x))|$$

\subsection{Recurrent Neural Network AutoEncoder (RNNAE)}
\subsubsection{Background}
The third architecture we tested is a Recurrent Neural Network AutoEncoder (RNNAE) which combines  aspects of AutoEncoders and recurrent neural networks (RNNs)~\cite{pascanu2013construct}. The rationale behind it is the observation that the traces in a provenance databases (e.g. the one used in the evaluation study presented below) are 
sequences of events. Therefore, we employ RNNs as an encoding engine for sequential data: The data is encoded into a compressed representation and then decoded back, roughly into its original form.

\subsubsection{Architecture}
The architecture of the RNNAE is similar to the one described in section~\ref{autoencoder}. The main difference resides in the nature of the elementary units (neurons) of the net. Here the nodes of the neural networks have the recurrent type. At each time step, the RNN unit receives an input vector \( x_t \), which represents the input data at that time step. 
Then, the RNNAE maintains a hidden state vector \( h_t \) that represents the network's internal memory. The hidden state captures information from previous time steps and influences the network's behavior at the current time step. The hidden state is computed based on the current input and the previous hidden state, often using a nonlinear activation function such as the hyperbolic tangent (tanh) or the rectified linear unit (ReLU).
    \[ h_t = f(W_{hx}x_t + W_{hh}h_{t-1} + b_h) \]
  where \( f \) is the activation function, \( W_{hx} \) and \( W_{hh} \) are the weight matrices, and \( b_h \) is the bias vector.
    
The loss function as-well as the anomaly scores are calculated in the same manner as the baseline AutoEncoder.

\subsection{Long Short-Term Memory AutoEncoder (LSTMAE)}
\subsubsection{Background}
The Long Short-Term Memory AutoEncoder (LSTMAE) is a special case of the RNNAE. Here we utilize LSTM units that address the vanishing gradient problem by introducing gating mechanisms for controlling the flow of information through the neural unit~\cite{hochreiter1997long}. They have separate gates for controlling the flow of information (input gate), forgetting information (forget gate), and outputting information (output gate), allowing them to capture long-term dependencies more effectively. \\
Except the class of the neurons, we kept the same architecture as the baseline AE as well as the evaluation and scoring protocols.

\subsection{Gated Recurrent Units AutoEncoder (GRUAE)}
\subsubsection{Background}
The Gated Recurrent Units AutoEncoder (GRUAE) is also a special case of the RNNAE. We used GRU units that are similar to LSTM units but have a simpler architecture with fewer parameters~\cite{chung2014empirical}. They combine the input and forget gates into a single update gate and use a single gate to control the output, making them computationally more efficient while still being effective for capturing long-range dependencies.\\
The architecture of the neural networks also remains the same as the AE model, except the class of the neural units.

\subsection{Attention-based AutoEncoder (ATAE)}
\subsubsection{Background}
Attention mechanism~\cite{bahdanau2016neural} is a concept in machine learning, particularly in deep learning models, that allows the model to focus on specific parts of the input data 
when making predictions or generating outputs. Originally inspired by human attention mechanisms, it has become a fundamental component in many state-of-the-art models, particularly in natural language processing (NLP) tasks such as machine translation, text summarizing, and question answering~\cite{vaswani2023attention}.
\subsubsection{The Need for Attention}
Traditional neural network architectures, such as recurrent neural networks (RNNs) and convolution neural networks (CNNs), process sequential data by treating all elements of the sequence equally. However, this approach may not be optimal for tasks involving long sequences or where certain elements are more important than others. The attention mechanism addresses this limitation by enabling models to selectively attend to relevant parts of the input, leading to more accurate and contextually rich predictions.
\subsubsection{Components of the Attention Mechanism}
The attention mechanism consists of several key components:
\subsubsection*{Query, Key, and Value:} In the attention mechanism, each element in the input sequence is associated with a key and a value, while the current state of the model (the query) determines which elements to attend to.
Mathematically, we represent the query, key, and value vectors as $q$, $k$ and $v$ respectively. These vectors are typically obtained by projecting the input sequence embeddings onto lower-dimensional spaces through learned weight matrices.
\subsubsection*{Attention Scores:} Attention scores quantify the relevance of each key to the query. A common way to compute attention scores is by measuring the similarity between the query and each key.
Mathematically, the attention score $e_{i}$ for the 
$i-th$ key is often computed using a similarity function $f$ applied to the query and key: $e_{i}=f(q,k_{i})$. Common similarity functions include dot-product attention, additive attention, and multiplicative attention.
\subsubsection*{Attention Weights:} The attention scores are normalized to obtain attention weights, which determine the importance of each input element. Higher attention weights indicate greater relevance. Mathematically, the attention weight is calculated as: $\alpha{i}=\frac{exp(e_{i})}{\Sigma_{j=1}^{N}exp(e_{j})}$. Here, $N$ is the number of keys in the input sequence.
\subsubsection*{Context Vector:} A context vector is computed as a weighted sum of the input elements, with the attention weights serving as the weights. This context vector captures the attended information from the input sequence.
\subsubsection*{Integration with Model:} The context vector is integrated into the model's computations to generate the output or make predictions. In sequence-to-sequence models, it may be used as input to the decoder to produce the next token in the output sequence.
\subsubsection{Architecture}
AutoEncoder architectures augmented with attention mechanisms can be particularly effective for anomaly detection.
The attention mechanism helps the model to reconstruct normal data more accurately by focusing on relevant features, making it more sensitive to deviations from normal behavior. In the previous AutoEncoder architectures, the entire input is compressed into a fixed-size latent representation, which may not effectively capture the most salient features of the input data, especially in complex or high-dimensional datasets. Attention mechanisms, inspired by human visual attention, allow the model to selectively focus on important parts of the input while disregarding irrelevant information.

\subsubsection*{Encoder with Attention Mechanism: }The encoder of the proposed ATAE Attention AutoEncoder processes the input data while dynamically attending to different parts of the input. Instead of generating a single fixed-size latent representation, the encoder produces a set of attention weights that indicate the importance of each input element. These attention weights are then used to compute a weighted sum of the input features, resulting in a context vector that captures the relevant information for encoding.

\subsubsection*{Decoder with Attention Mechanism:} Similarly, the decoder of the ATAE Attention AutoEncoder employs attention mechanisms to selectively attend to different parts of the context vector generated by the encoder. This allows the decoder to reconstruct the input by focusing on the most relevant features, leading to improved reconstruction quality.

We've implemented ATAE using the AttentionLayer class, which implements the attention mechanism\footnote{https://keras.io/api/layers/attention\_layers/attention/}. This layer computes attention weights using a trainable weight matrix and applies those weights to the input.
The encoder part consists of a dense layer, followed by the attention layer. The decoder part consists of another dense layer.
To include an anomaly score, we compute as in the previous architectures a reconstruction error between the input and the reconstructed output. Anomalies are typically instances that have higher reconstruction errors compared to normal data. 

\section{Experimental settings, results and analysis}

Below, we present the datasets we used, our experimental protocol, and the outcome of various simulations together with our interpretation thereof. The databases, codes, and reproducibility guidelines are given in the project's repository\footnote{https://github.com/ae-apt/AE-APT}.

\subsection{Datasets}\label{datasets}
The data used in this paper comes from the Defense Advanced Research Projects Agency (DARPA)’s \verb|Transparent| \verb|Computing TC| program~\cite{darpa}. The aim of this program is to provide transparent provenance data of system activities and component interactions across different operating systems (OS) and spanning all layers of software abstractions. Specifically, the datasets include system-level data, background activities, and system operations recorded while APT-style attacks are being carried out on the underlying systems. Preserving the provenance of all system elements allows for tracking the interactions and dependencies among components. Such an interdependent view of system operations is helpful for detecting activities that are individually legitimate or benign but collectively might indicate abnormal behavior or malicious intent.

\begin{figure}
    \centering
    \includegraphics[width=0.7\linewidth]{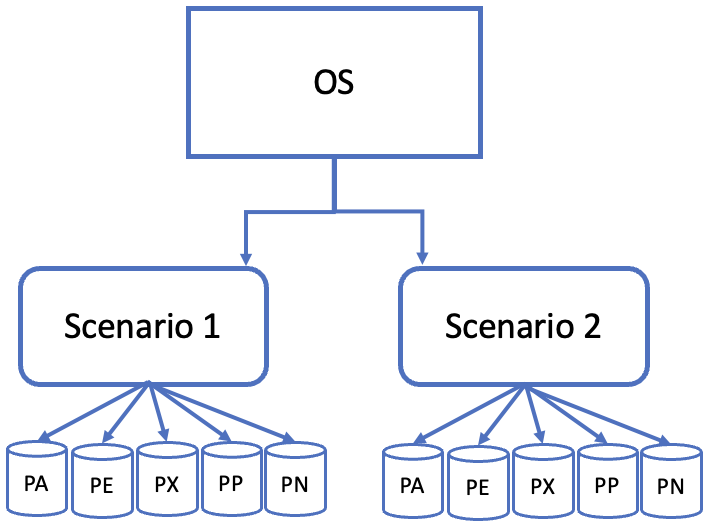}
    \caption{Organization of the DARPA's TC datasets. Each OS undergoes two attack scenarios, each of which contains five datasets. With four OS (BSD, Windows, Linux, Android), two attack scenarios, and five aspects (PE, PX, PP, PN, PA), a total of forty individual datasets are composed.}
    \label{fig:enter-label}
\end{figure}

Here we specifically employ DARPA’s data that has undergone processing conducted by the \verb|ADAPT| (Automatic Detection of Advanced Persistent Threats) project’s ingester~\cite{berrada_2019,BerradaCBMMTW20,Benabderrahmane21}. The records come from four different source OS, namely Android (called in the TC program Clearscope), Linux (called Trace), BSD (called Cadets), and Windows (called Fivedirections or 5dir). For each system, the data comes from two separate attack scenarios: scenario 1 and scenario 2, called Pandex (Engagement E1) and Bovia (Engagement E2), respectively. The processing includes ingesting provenance graph data into a graph database as well as additional data integration and deduplication steps. The final data includes a number of Boolean-valued datasets
, with each representing an aspect of the behavior of system processes (see Table~\ref{dataexample}). Each row in such a dataset 
is a data point representing a single process run on the respective OS. It is expressed as a Boolean vector whereby a value of 1 in a vector cell indicates the corresponding attribute applies to that process. For instance, in Table~\ref{dataexample} the process with id \verb|ee27fff2-a0fd-1f516db3d35f| has the following sequence of events: \verb|</usr/sbin/avahi-autoipd|, \verb|216.73.87.152|, \\ \verb|EVENT_OPEN, EVENT_CONNECT, ...>|. Specifically, the relevant
datasets are interpreted as follows:

\begin{itemize}
    \item \verb|ProcessEvent| (PE): Its attributes are event types performed by the processes. A value of 1 in \verb|process[i]| means the process has performed at least one event of type $i$.
    \item \verb|ProcessExec| (PX): The attributes are executable names that are used to start the processes.
    \item \verb|ProcessParent| (PP): Its attributes are executable names that are used to start the parents of the processes.
    \item \verb|ProcessNetflow| (PN): The attributes here represent IP addresses and port names that have been accessed by the processes.
    \item \verb|ProcessAll| (PA): This dataset is described by the disjoint union of all attribute sets from the previous datasets.
\end{itemize}
Overall, with two attack scenarios (Pandex, Bovia), four OS (BSD, Windows, Linux, Android) and five aspects (PE, PX, PP, PN, PA), a total of forty individual datasets are composed (Figure \ref{fig:enter-label}). 
They are described in Table~\ref{datatable} whereby the last column provides the number of attacks in each dataset. The substantially imbalanced nature of the datasets is clearly seen here.
\begin{table*}
\vspace{-6 em}
\centering
\small
\resizebox{0.99\textwidth}{!}{
\begin{tabular}{|l|l||l|l|l|l|l|l|l|l|}
\hline & Scenario & Size& $PE$   & $PX$  & $PP$  & $PN$     & $PA$  & $nb\_attacks$    & $\%\frac{nb\_attacks}{nb\_processes}$     \\ \hline \hline
BSD    & 1 &288 MB &76903 / 29  & 76698 / 107  & 76455 / 24  & 31 / 136  & 76903 / 296 & 13&0.02\\  
    & 2 &1.27 GB &224624 / 31  &224246 / 135  & 223780 / 37  & 42888 / 62 &  224624 / 265      & 11&0.004\\ \hline
Windows & 1 &743 MB & 17569 / 22    &  17552 / 215  &   14007 / 77        &   92 / 13963      & 17569 / 14431& 8&0.04\\  
   & 2 &9.53 GB& 11151 / 30    &  11077 / 388  & 10922 / 84  & 329 / 125      &  11151 / 606    &8&0.07\\ \hline
Linux  & 1 &2858 MB &247160 / 24 & 186726 / 154 & 173211 / 40 & 3125 / 81 & 247160 / 299  &25&0.01\\
    & 2 &25.9 GB &282087 / 25 & 271088 / 140 & 263730 / 45 &6589 / 6225 &  282104 / 6435      &46&0.01\\ \hline
Android& 1 &2688 MB&102 / 21     &102 / 42&0 / 0&8 / 17& 102 / 80&9&8.8\\
&2 &10.9 GB&12106 / 27     &12106 / 44&24 / 11&4550 / 213&12106 / 295 &13&0.10\\ \hline
\end{tabular}
}

\caption{Experimental datasets of DARPA's TC program used in our study. A dataset entry (columns 4 to 8) is described by a number of rows (processes) / number of columns (attributes). For instance, with ProcessAll (PA) obtained from the second scenario using Linux, the dataset has 282104 rows and 6435 attributes with 46 APT attacks (0.01\%). }
 \label{datatable}
\end{table*}

\begin{table}
\scriptsize
\begin{tabular}{|l|c|c|c|c|c|c|c|}
\hline
                                           & {\rotatebox{90}{/usr/sbin/avahi-autoipd}} & {\rotatebox{90}{216.73.87.152}}   & {\rotatebox{90}{EVENT\_OPEN} }    & {\rotatebox{90}{EVENT\_EXECUTE}}  & {\rotatebox{90}{EVENT\_CONNECT} } & {\rotatebox{90}{EVENT\_SENDMSG} } & \multicolumn{1}{l|}{...} \\ \hline
{ee27fff2-a0fd-1f516db3d35f} & 1                                & 1                        & 1                        & 0                        & 1                        & 0                        & ...                      \\ \hline
{b2e7e930-8f25-4242a52c5d72} & 0                                & 1                        & 0                        & 1                        & 1                        & 1                        & ...                      \\ \hline
{07141a2a-832e-8a71ca767319} & 0                                & 0                        & 1                        & 1                        & 1                        & 1                        & ...                      \\ \hline
{b4be70a9-98ac-81b0042dbecb} & 1                                & 0                        & 1                        & 1                        & 0                        & 0                        & ...                      \\ \hline
{2bc3b5c6-9110-076710a13038} & 0                                & 0                        & 0                        & 0                        & 0                        & 1                        & ...                      \\ \hline
{ad7716e0-8d59-5d45d1742211} & 1                                & 1                        & 0                        & 1                        & 0                        & 1                        & ...                      \\ \hline
...                                           & \multicolumn{1}{l|}{...}         & \multicolumn{1}{l|}{...} & \multicolumn{1}{l|}{...} & \multicolumn{1}{l|}{...} & \multicolumn{1}{l|}{...} & \multicolumn{1}{l|}{...} & \multicolumn{1}{l|}{...} \\ \hline
\end{tabular}
\caption{Example of a dataset. In each row, the boolean vector represents lists the features of the corresponding process. }
 \label{dataexample}
\end{table}

\begin{figure}[!htb]
     \centering
     \includegraphics[width=\linewidth]{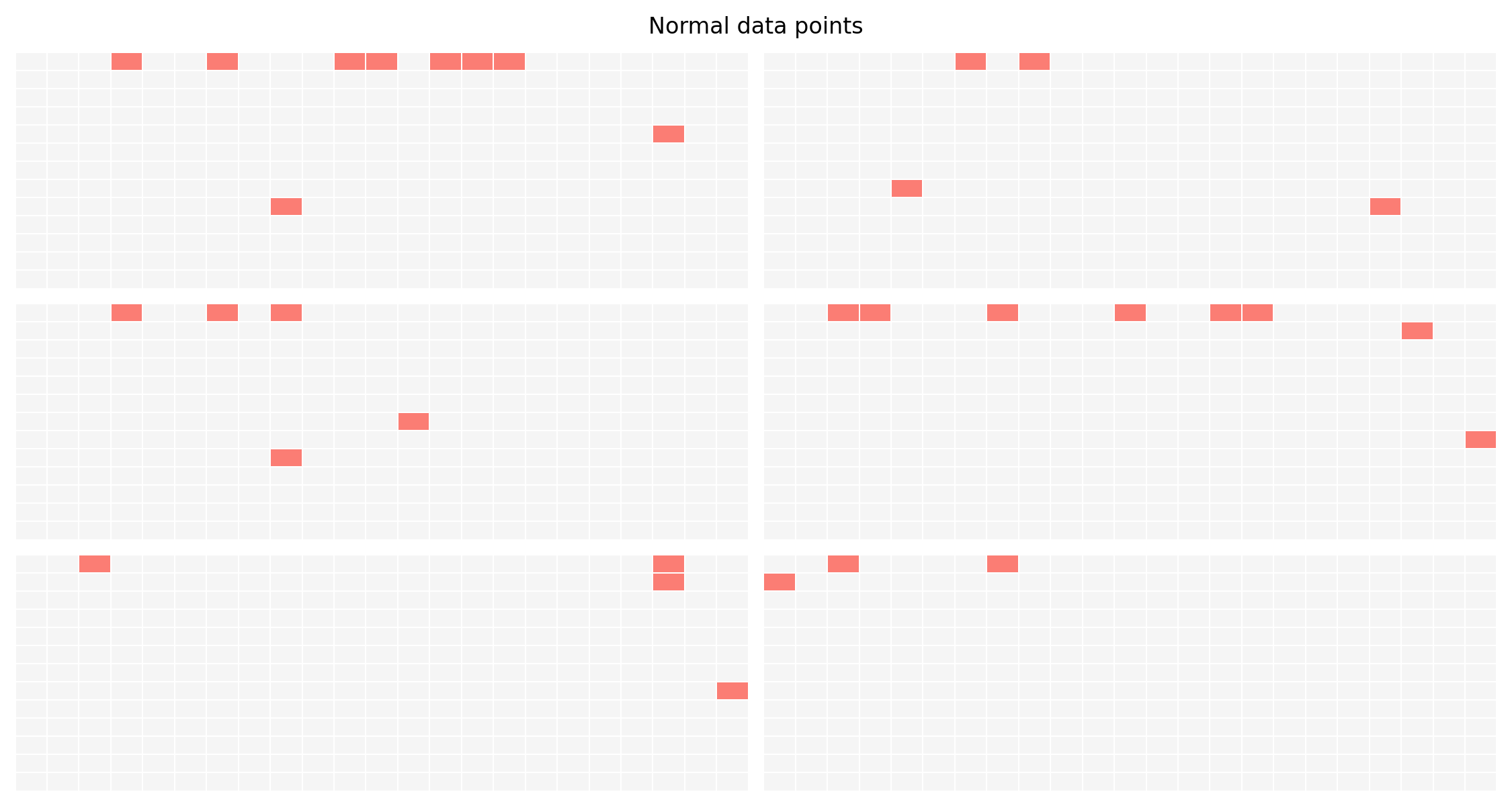}
     \caption{Visualization of 6 normal data points, sampled from the ProcessAll dataset of the Linux (Trace) system, Pandex scenario.}\label{Fig:normal_data}
     \hfill 
\includegraphics[width=\linewidth]{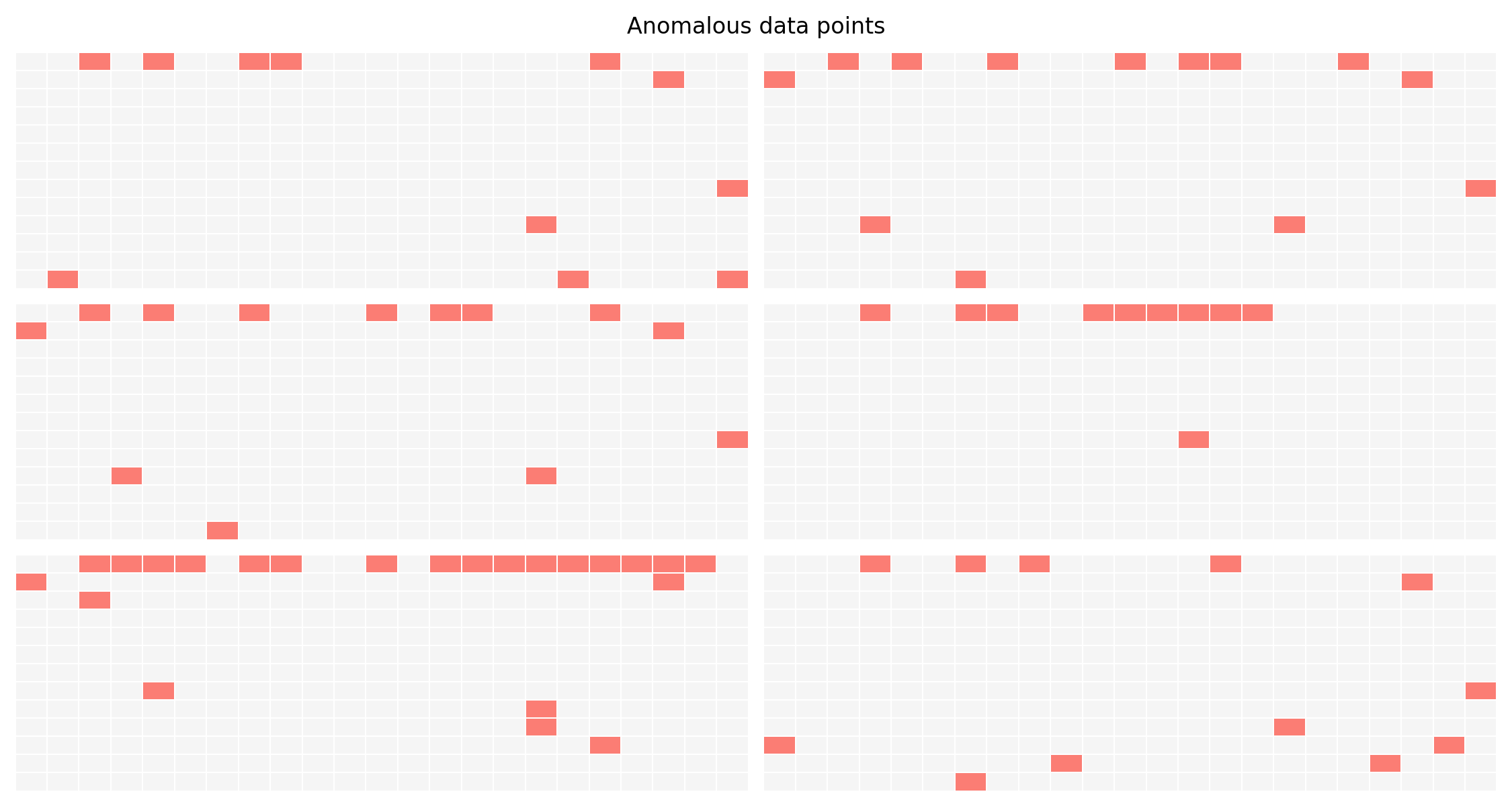}
     \caption{Visualization of 6 anomalous data points, sampled from the ProcessAll dataset of the Linux (Trace) system, Pandex scenario.}\label{Fig:anomalous_data}
\end{figure}

Figures~\ref{Fig:normal_data} and \ref{Fig:anomalous_data} illustrate 6-strong samples of normal and of anomalous data points, respectively. They all belong to the \verb|ProcessAll| context of the Linux system, Pandex scenario. Each subfigure 
represents a single data point whereby, for visualization purposes, the 299 entries of the binary vector are arranged into a $13 \times 23$ grid. 
Red rectangles (darker) represent the 1s while the gray ones (lighter) the 0s. The dataset is clearly very sparse. Moreover, with normal data points, the 1-valued features mostly rank in the 1st half of the vector and there are relatively few of them. In contrast, anomalous point vectors have comparatively larger sets of 1-valued features with many of them located in the 2nd half.

The rate of truly anomalous processes associated with APT-style attacks remains low, usually under 0.1\% of the data. A dataset might have 
many tens of 1000s of rows and up to a few thousands of columns where only few dozens, if not less than a dozen, of rows are, in actuality, relevant to malicious attacks.

\subsection{Computational environment}
The experiments were conducted on a machine running macOS 12.3.1 with an Apple M1 silicon chip, 8 GB RAM. The deep learning models were tested on every one of the 40 datasets.

\subsection{Evaluation metrics}

Since the anomalous and malicious processes account for only under 0.1\% of the data, accuracy would be a poor choice for evaluation metric. Indeed, an algorithm can reach a near 100\% accuracy by deeming all rows normal. Therefore, we focus on model's ability to identify attack-related processes instead of the correct labeling of the normal ones. 
This emphasizes the assessment of the relative anomaly status of a row, typically through a scoring function. By assigning higher scores to anomalous data points, a ranking of the processes w.r.t. their perceived anomalous status will be effectively produced. 

To that end, we selected the normalized discounted cumulative gain (nDCG)~\cite{jarvelin_2002} metric. It is typically used in Information Retrieval to evaluate competing methods based on their ability to retrieve highly relevant entities. The metric fits well the anomalous processes detection as it assesses the collective ranking of a set of target documents within the overall score-sorted list~\cite{BerradaCBMMTW20,Benabderrahmane21}. In our settings, the nDCG score of a ranking ranges from 0 to 1, with 1 being the best achievable score, i.e. in the ideal scenario where all anomalies are ranked at the top.

The first step in getting the nDCG score is computing the discounted cumulative gain (DCG) score. As a starting point, each entry in the ranking gets its relevance score reduced by a factor proportional to (the logarithm of) its rank. More formally, the DCG score of a ranking with $N$ entries is calculated as follows:
$$DCG = \sum_{i=1}^{N} \frac{rel_i}{\log_2{(i+1)}}$$
where $rel_i$ is the relevance score of the $i$\textsuperscript{th} entry in the ranking.
Furthermore, to factor in the total number of entries in the ranking, the DCG score is normalized by the ideal DCG score (iDCG) which is the DCG score corresponding to all relevant entries being ranked at the very top of the list: $$nDCG = \frac{DCG}{iDCG}.$$

In all our experiments, each process of a dataset is assigned an anomaly score which is the basis of a subsequent ranking among those. If the dataset has $k$ anomalous data points, the best possible score is obtained when their anomaly scores are invariably higher than any of the normal process scores. This results in anomalies being ranked at the top of the list (the precise order among them is immaterial). Therefore, the best nDCG score is achieved when all anomalous processes are ranked at the top. This approach ensures that the evaluation focuses on the model's ability to identify and rank attack-related processes accurately.P
This metric is particularly useful in scenarios where the primary goal is to detect and prioritize anomalies rather than just classifying data points correctly. It provides a more meaningful measure of performance in the context of anomaly detection for Advanced Persistent Threats (APTs).

\subsection{Results: Within AE-APT Comparison}

In the following paragraphs, the performances of the six AE-APT models we proposed above (Baseline AutoEncoder AE, Adversarial AutoEncoder AAE, Recurrent Neural Network AutoEncoder RNNAE, Long Short-Term Memory AutoEncoder LSTMAE, Gated Recurrent Unit AutoEncoder GRUAE, and Attention-based AutoEncoder ATAE) are discussed. They were evaluated on the datasets described in the previous section. %

In the present settings, for each experiment, the corresponding model is trained exclusively on data labeled as normal processes. The nDCG scores of all the experiments are summarized in Table~\ref{tab:ndcg}. Bold-printed values in each row represent the highest nDCG score reached by a model among all the experiments for a specific OS $\times$ attack scenario  combination, i.e. on any of the corresponding datasets. Such a value thus represents the best ranking quality a model can reach in that particular forensic configuration. The last column of the table represents the winner model for each pair of OS and attack scenario. In the current experiment batch, we set the aggregation rule in our ensemble learning to the maximal value of nDCGes. Albeit, arguably, the most intuitive choice, it is by far not the only one.

As a general trend, one might notice that the highest scores were primarily reached on the \verb|ProcessAll| and \verb|ProcessEvent| datasets. We tend to see this trend as most likely due to the rich structure of these datasets. During the training, such structure tends to yield a higher-quality mapping onto a reduced-dimension space. The scores reached on \verb|ProcessNetflow|, a very sparse family of datasets, seem to support such hypothesis: Indeed, across all systems and scenarios, these scores are substantially lower than the maximum. One potential reason for this could be the very high dimensionality but low volume of data within the corresponding datasets. As a result, the models fail in their task of  learning a meaningful latent space representation of normality. It is noteworthy that acceptable scores were also obtained with \verb|ProcessExec| and \verb|ProcessParent| for the second attack scenario. 

Overall, the ATAE model has shown the best performances: It was the single top model in 6 out of 8 cases, with nDCG score ranging in $[0.7, 0.92]$. 
The best score of 0.92 was achieved on data recorded on the Linux OS (Trace), with the second attack scenario (Bovia) and the \verb|ProcessAll| format. It is closely followed by the score reached by the AAE model on \verb|ProcessAll| data of the BSD system with the same Bovia scenario (0.91). 

\begin{figure*}[h!]

%\vspace{-4 em}
\begin{minipage}{.52\textwidth}
    \subfloat{\includegraphics[width=\textwidth]{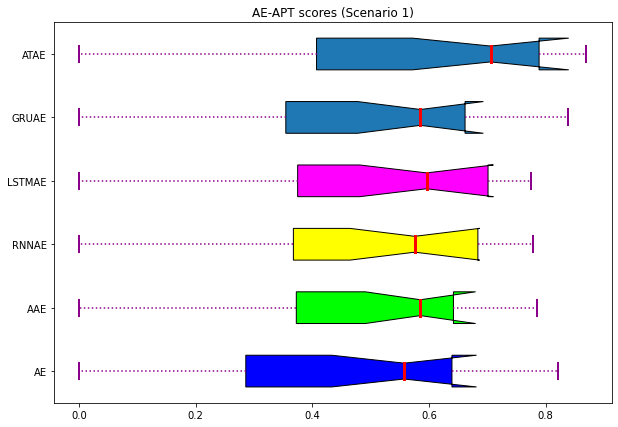}}\label{fig:sub_a}
\end{minipage}
\hfill    
\begin{minipage}{.52\textwidth}
    \subfloat{\includegraphics[width=\textwidth]{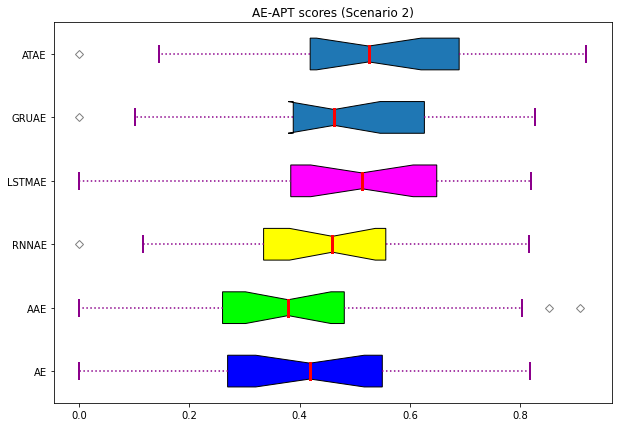}}\label{fig:sub_b}
\end{minipage}
      \caption{Standalone and Stacking Models with all OSs and across 2 attack scenarios (left to right). X-axis: the range of the nDCG scores. Y-axis:  Standalone AutoEncoder model. }\label{fig:ensemble}
\end{figure*}

Figure~\ref{fig:ensemble} illustrates the range of the nDCG scores by standalone models within  AE-APT in the form of box-plots across all operating systems and attack scenarios. The x-axis represents the range of nDCG scores, while the y-axis lists the different standalone AutoEncoder models. The maximal values of individual learners provide the basis for the selection of a winner model w.r.t. to a scenario. For instance, observe that ATAE has maximal nDCG values ranging over 0.9. It consistently outperforms the other models, achieving the highest nDCG scores. This indicates that the ATAE model is the most effective at ranking anomalous processes at the top of the list, reflecting its superior ability to detect APTs. The median nDCG scores for ATAE are notably higher than those of other models, demonstrating its robustness and accuracy. In the same Figure, the box plots for the other models (AE, AAE, RNNAE, LSTMAE, GRUAE) reveal significant variability in their performance, with some models performing well in certain scenarios but not consistently across all datasets. This variability highlights the challenges in anomaly detection and the importance of selecting an appropriate model for specific contexts. For instance, the baseline AE and the adversarial one, AAE, proved to be inter-competitive: Neither model consistently outperforms the other nor performs consistently well across the datasets of the same OS or the same-type dataset across different systems. We tend to see this as mostly due to the large variability among the datasets in terms of dimensionality, volume and proportion of anomalies. Beside the lack of clear superiority, both models performed  very well on most datasets: They achieved maximal nDCG scores above 0.7 in many Pandex cases and even above 0.8 in the Bovia ones (see also the corresponding scores in Table~\ref{tab:ndcg}). 
The recurrent models (RNNAE, LSTMAE and GRUAE) also proved competitive to each other. Their higher scores were often obtained on large datasets where the number of features is high, hence they make for longer event sequences. The maximal values in the box-plots reflecting their best performances, once more, are very close. They consistently range above 0.75 and 0.8 for the first and for the second scenario, respectively. Overall, Figure~\ref{fig:ensemble} underscores the effectiveness of the attention mechanism in enhancing anomaly detection capabilities, as evidenced by the high performance of the ATAE model. It also illustrates the competitive performance of recurrent models like RNNAE, LSTMAE, and GRUAE, which perform well on larger datasets with longer event sequences.

\begin{figure*}
%\vspace{-3 em}
\begin{minipage}{.52\textwidth}
    \subfloat{\includegraphics[width=\textwidth]{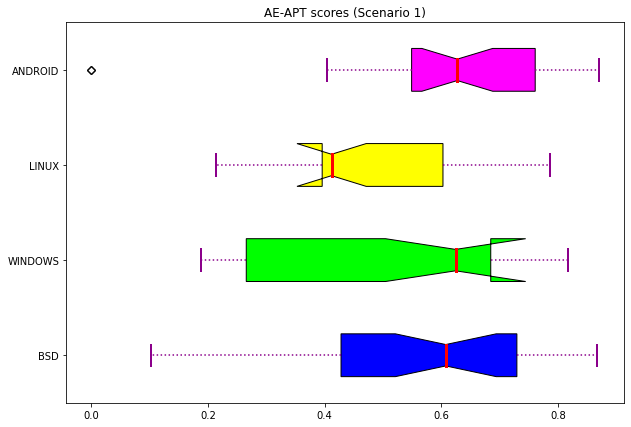}}\label{fig:sub_a2}
\end{minipage}
\hfill    
\begin{minipage}{.52\textwidth}
    \subfloat{\includegraphics[width=\textwidth]{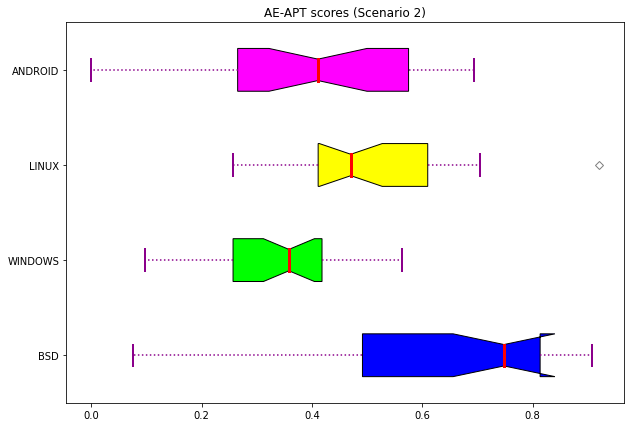}}\label{fig:sub_b2}
\end{minipage}
      \caption{Performance of AE-APT when considering the different models per operating system. X-axis: the range of the nDCG scores. Y-axis: OS.}\label{fig:os-wise}
\end{figure*}

An orthogonal perspective is shown in Figure~\ref{fig:os-wise}: The drawings depict the scopes of the maximal nDCG scores OS-wise. The x-axis represents the range of nDCG scores, while the y-axis shows the different operating systems.
The box plots provide a clue about how difficult APT detection is in the individual OS cases, as reflected in the maximal value of each the box-plots. 
It is noteworthy that the eight extreme values represent the top nDCG scores after models' aggregation. They are also to be found in the last column of Table~\ref{tab:ndcg}. The figure also reveals several key insights. For instance, concerning Linux (Trace): The family of AutoEncoder models achieve high nDCG scores, indicating that they perform exceptionally well in detecting anomalies within this OS, with a top nDCG scores close to 0.92. Regarding BSD (Cadets): The  models also perform strongly on BSD datasets, with high nDCG scores, particularly for the ProcessAll and ProcessEvent datasets. The highest nDCG score reaches approximately 0.91, showing robust anomaly detection capabilities. Concerning Windows (5dir): The performance shows more variability, with lower nDCG scores compared to Linux and BSD. The top nDCG scores for Windows are around 0.82, indicating that while the AutoEncoder models are effective, they may face more challenges in this OS. Finally, within Android (Clearscope): The results are mixed. While some AutoEncoder models achieve relatively high nDCG scores, there is a noticeable drop in performance for certain datasets. The best nDCG scores for Android are around 0.87, highlighting that the AutoEncoder models can still perform well.

Below, we discuss a second part of our experimental study in which the neural models in AE-APT are compared with some state-of-the-art methods. For simplicity, we will denote our elected winner models (see last column in Table~\ref{tab:ndcg}) simply as AE-APT and use the aforementioned maximal nDCG scores as a collective performance indicator for each case.

\begin{table*}[t!]
\centering
\scriptsize
\vspace{-3 em}%
%\Rotatebox{90}
{%
\begin{tabular}{lllcccccr}
\hline
\textbf{OS}&\textbf{Attack} &\textbf{Model  }                                                                                                             & 
\textbf{ProcessAll} & \textbf{ProcessEvent} & \textbf{ProcessExec} & \textbf{ProcessParent} & \textbf{ProcessNetflow} & \textbf{Max }\\ 
\textbf{}&\textbf{Scenario} &\textbf{  }                                                                                                             & 
\textbf{} & \textbf{} & \textbf{} & \textbf{} & \textbf{} & \textbf{ Aggregationing}\\ \hline

\multirow{4}{*}{\textbf{\begin{tabular}[c]{@{}l@{}}Cadets\\ (BSD)\end{tabular}}}         & \multirow{2}{*}{\textbf{Pandex}} & AE             & \textbf{0.8215  }     & 0.5791         & 0.8165        & 0.5358          & 0.1165        &   \\
                                                                                         &                                  & AAE  & \textbf{0.7268}       & 0.6394         & 0.6276        & 0.5822          & 0.1175     &      \\ 
                                                       
                                                                                         &                                  & RNNAE& \textbf{0.7013 }  &   0.6154      &    0.4401     &    0.4239   &   0.1019  &   \\   
                                                       
                                                                                         &                                  & LSTMAE&  0.6982  &   \textbf{0.7453 }      &     0.5732    &    0.5998   &    0.1243  &  ATAE\\     
                                                       
                                                                                         &                                  & GRUAE&    \textbf{0.8380}&  0.7304        &  0.4092       &   0.5423    &  0.1260  & $\approx$0.87   \\   
                                                       
                                                                                         &                                  & ATAE& \textbf{0.8676}   &   0.7723       &   0.7923      &    0.6645   &    0.1203   & \\                          \cline{2-8} 
                                                                                         & \multirow{2}{*}{\textbf{Bovia}}  & AE             & 0.8077       & 0.4823         & \textbf{0.8168 }       & 0.7992          & 0.1519     &      \\
                                                                                         &                                  & AAE & \textbf{0.9079 }      & 0.4848         & 0.8524        & 0.8035          & 0.0760      &     \\ 
                                                             
                                                                                         &                                  & RNNAE& 0.7144   &   0.5113       &    \textbf{0.8155}     &0.7765       &     0.1165 &  \\   
                                                       
                                                                                         &                                  & LSTMAE&   \textbf{0.8200} &    0.5344      &    0.8066     &    0.7054   &0.1356    & AAE   \\     
                                                       
                                                                                         &                                  & GRUAE&      0.7647       &    0.5444    &   \textbf{0.8276}   &  0.7315  &0.1023 &$\approx$0.91    \\   
                                                       
                                                                                         &                                  & ATAE&\textbf{0.8938}    &     0.5377     &    0.8345     &    0.7903   &     0.1456&   \\                                         \hline
\multirow{4}{*}{\textbf{\begin{tabular}[c]{@{}l@{}}5dir\\ (Windows)\end{tabular}}}       & \multirow{2}{*}{\textbf{Pandex}} & AE             & 0.5931       & \textbf{0.7086  }       & 0.2272        & 0.1970          & 0.6289           &\\
                                                                                         &                                  & AAE & 0.5989       & \textbf{0.6676 }        & 0.2882        & 0.1871          & 0.6495     &     \\ 
                                                             
                                                                                         &                                  & RNNAE& 0.6834   &    \textbf{0.6851  }    &     0.2799    &   0.2618    &   0.6864 &    \\   
                                                       
                                                                                         &                                  & LSTMAE&  0.7104 &    \textbf{0.7746}      &    0.2765     &   0.1988    &  0.6588&     ATAE \\     
                                                       
                                                                                         &                                  & GRUAE&    0.6390&  \textbf{0.7519 }       &     0.2011    &  0.2033     &    0.6200&$\approx$0.82    \\   
                                                       
                                                                                         &                                  & ATAE&   0.7876 &   \textbf{0.8173 }      &   0.2899      &    0.2525   &    0.6713 &   \\                                         \cline{2-8} 
                                                                                         & \multirow{2}{*}{\textbf{Bovia}}  & AE             & 0.4072       & 0.2639         & 0.3256        & \textbf{0.4303}          & 0.0973      &     \\
                                                                                         &                                  & AAE  &   0.3952         & 0.2514         & 0.3218        & \textbf{0.4200 }         & 0.1298     &      \\ 
                                                                                         
                                                                                         &                                  & RNNAE&   0.4234 &   0.2544       &   0.3316      &    \textbf{0.4343 }  &     0.1709&   \\   
                                                       
                                                                                         &                                  & LSTMAE& \textbf{0.5635}   &    0.3345      &  0.3351       &  0.4123     &     0.1322&   LSTMAE\\     
                                                       
                                                                                         &                                  & GRUAE&  0.3811  &    \textbf{0.4819 }     & 0.2908        &  0.4019     &   0.1190  &$\approx$0.56   \\   
                                                       
                                                                                         &                                  & ATAE&   \textbf{0.5566} &   0.4123       &  0.3898       &    0.5009   &   0.1798 &    \\             \hline
\multirow{4}{*}{\textbf{\begin{tabular}[c]{@{}l@{}}Trace\\ (Linux)\end{tabular}}}        & \multirow{2}{*}{\textbf{Pandex}} & AE             & \textbf{0.6064   }    & 0.4204         & 0.3013        & 0.2397          & 0.3953          \\
                                                                                         &                                  & AAE  & \textbf{0.7711 }      & 0.4916         & 0.4080        & 0.2278          & 0.4006      &     \\
                                                                 
                                                                                         &                                  & RNNAE&   0.5901 &   \textbf{0.6301}       &   0.4126      &    0.2139   &    0.3967 &   \\   
                                                       
                                                                                         &                                  & LSTMAE&  \textbf{0.6213}  &   0.5918       &   0.4098      &   0.2312    &  0.4076 & ATAE    \\     
                                                       
                                                                                         &                                  & GRUAE&  0.6209  &    \textbf{0.6259}      &   0.3987      &   0.2219    &  0.4109 &$\approx$0.79     \\   
                                                       
                                                                                         &                                  & ATAE&   \textbf{0.7856} &   0.7401       &    0.4565     &    0.2312   &   0.4463     \\                                     \cline{2-8} 
                                                                                         & \multirow{2}{*}{\textbf{Bovia}}  & AE             & \textbf{0.7054 }      & 0.4761         & 0.4871        & 0.2564          & 0.3725      &     \\
                                                                                         &                                  & AAE & \textbf{0.6494 }      & 0.4234         & 0.4796        & 0.2850          & 0.3633      &     \\ 
                                                                                             
                                                                                         &                                  & RNNAE&  \textbf{0.6614}  &    0.4848      &   0.4845      &    0.3356   &  0.3908     & \\   
                                                       
                                                                                         &                                  & LSTMAE&  0.6703  &   \textbf {0.6878    }  &   0.4913      &    0.4453   &    0.4119& ATAE   \\     
                                                       
                                                                                         &                                  & GRUAE&     \textbf{0.6419}        &  0.4645       &  0.4606     &    0.4108    &
                                                     0.3908                                    &$\approx$0.92\\   
                                                       
                                                                                         &                                  & ATAE&   \textbf{0.9198} &   0.6708       &   0.5134      &     0.4478  &    0.4213 &   \\         
                                                                                        \hline
\multirow{4}{*}{\textbf{\begin{tabular}[c]{@{}l@{}}Clearscope\\ (Android)\end{tabular}}} & \multirow{2}{*}{\textbf{Pandex}} & AE             &\textbf{ 0.7815 }      & 0.6708         & 0.4033        & NA              & 0.6014           \\
                                                                                         &                                  & AAE  & \textbf{0.7857 }      & 0.5484         & 0.5885        & NA              & 0.6284      &     \\
                                                                                             
                                                                                         &                                  & RNNAE&   0.7609 &   \textbf{0.7791 }      &    0.5612     &    NA   &   0.6134 &    \\   
                                                       
                                                                                         &                                  & LSTMAE& 0.7509   &       \textbf{0.7656}   &  0.5617       &   NA    &    0.6313 &ATAE   \\     
                                                       
                                                                                         &                                  & GRUAE&   \textbf{0.7609} &        0.7517  &    0.5508     &   NA    &   0.6262&$\approx$0.87     \\   
                                                       
                                                                                         &                                  & ATAE&   0.8301 & \textbf{0.8698}         &   0.7813      &    NA   &   0.6613 &    \\         \cline{2-8} 
                                                                                         & \multirow{2}{*}{\textbf{Bovia}}  & AE             & 0.2712       & \textbf{0.6357 }        & 0.5212        & NA              & 0.3904      &     \\
                                                                                         &                                  & AAE  & 0.2949       & 0.2631         & \textbf{0.3952 }       & NA              & 0.1681     &      \\ 
                                                                                             
                                                                                         &                                  & RNNAE&   0.5456&   \textbf{0.5879 }     &    0.5309     &    NA   &   0.3787 &    \\   
                                                       
                                                                                         &                                  & LSTMAE& 0.5607   &   \textbf{0.6413 }      &    0.5798     &   NA    &   0.4001 &ATAE    \\     
                                                       
                                                                                         &                                  & GRUAE&   \textbf{0.6495} &   0.6209       &   0.4861      &    NA   &   0.3967 & $\approx$0.70   \\   
                                                       
                                                                                         &                                  & ATAE&   \textbf{0.6930} &   0.6879       &   0.5373      &    NA   &  0.4215 &     \\         \hline
\end{tabular}
}
\caption{nDCG scores of AE-APT AutoEncoder architectures on all available datasets. AE: Baseline AutoEncoder. AAE: Adversarial AutoEncoder. RNNAE: Recurrent AutoEncoder. LSTMAE: Long Short-Term Memory AutoEncoder. GRUAE: Gated Recurrent Unit AutoEncoder. ATAE: Attention AutoEncoder. NA: data not available. Bold values represent the max nDCG for each OS $\times$ attack scenario $\times$ dataset. The last column represents the elected winner approach for each forensic configuration.}
\label{tab:ndcg}
\end{table*}

%%%%%%%%%%%%%%%%%%%%%%%%%

\subsection{Results: Comparison to state-of-the-art methods}

Our neural models were benchmarked against a number of competing methods for APT-oriented AD (VF-ARM, VR-ARM, AVF, FPOF, OD and OC3). These were evaluated on the same DARPA TC datasets in a couple of previously studies~\cite{BerradaCBMMTW20,Benabderrahmane21}, hence they provide an appropriate baseline for performance assessment. We therefore kept the same experimental framework as above, i.e. the 40 datasets reflecting both attack scenarios on four different OS with five data sources each.

\begin{figure*}
\vspace{-6 em}%
  \centering
\includegraphics[width=\textwidth]{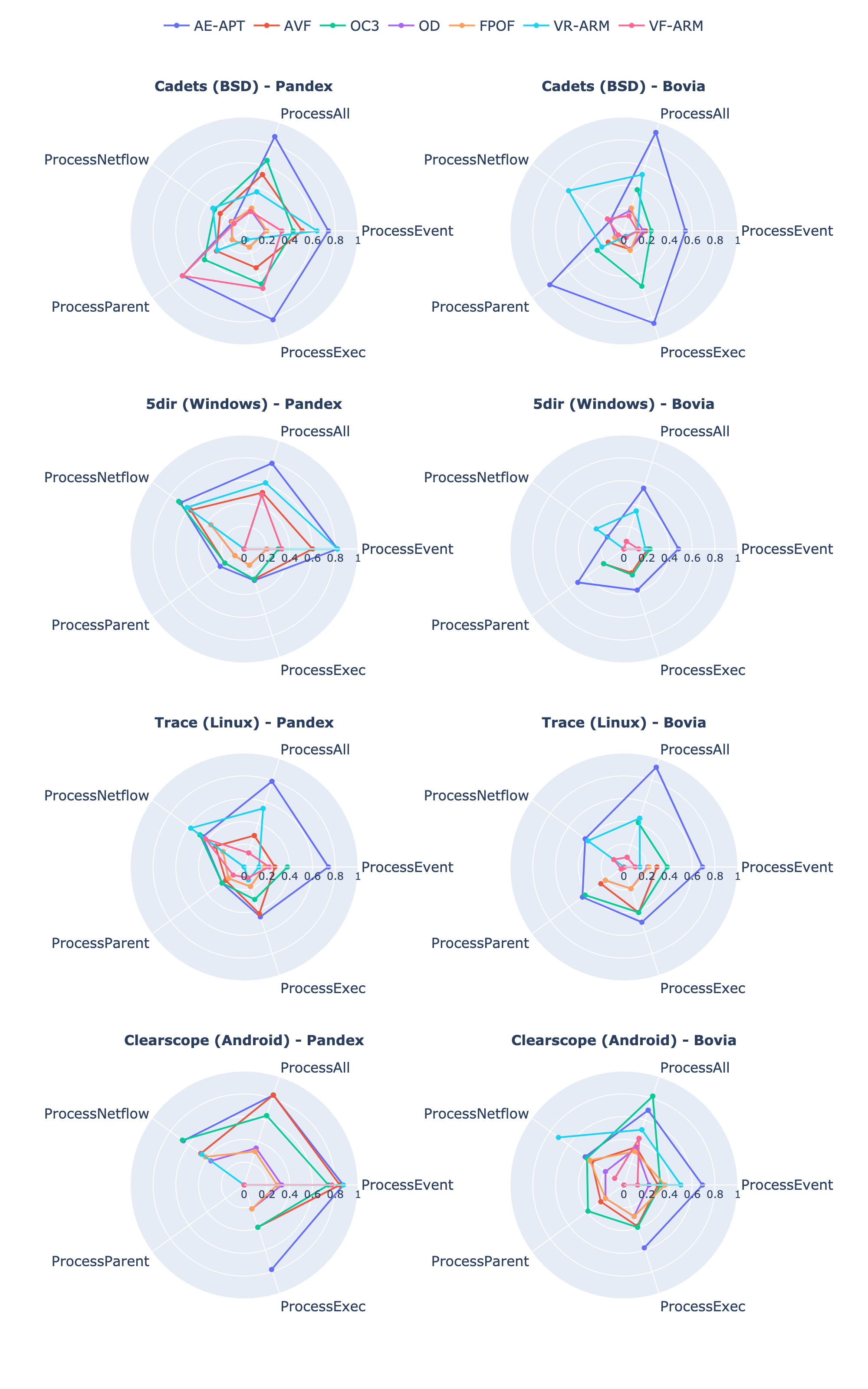}
\vspace{-6 em}%
  \caption{nDCG scores of AE-APT versus compared algorithms using BSD, Windows, Linux and Android systems, with two attack scenarios.}
  \label{fig:cadet-pandex}
\end{figure*}

Figure~\ref{fig:cadet-pandex} summarizes the outcome of our performance comparison in a radar plot.
Overall,  AE-APT outclassed the baseline algorithms. For the datasets recorded on the BSD system, Pandex scenario (upper left corner), in all columns except for \verb|ProcessNetflow|, the best nDCG scores are achieved by AE-APT. The highest score of 0.87 was achieved on \verb|ProcessAll| by the attention-based model of AE-APT (ATAE). The same pattern repeats for the Bovia scenario of the that OS (upper right corner), where AE-APT outperformed the competitors by a significant margin, except for the \verb|ProcessNetflow| case where the rare association rule mining method VR-ARM reached 0.60. The highest nDCG score, 0.91,  was obtained by our AAE model. 
For the datasets reflecting the Pandex scenario on Windows, the top performing methods were our new model ATAE and VR-ARM, both reaching the highest nDCG score of 0.82. Compared to the first scenario, all methods performed worse on the second one with the same OS. Nevertheless, the models in AE-APT still generally performed either on par with or better than the six baselines. Thus, LSTMAE reached a maximal nDCG of 0.56. 
Higher scores have also been obtained by AE-APT on the Linux datasets. Here, the top nDCG score is about 0.79 for Pandex and 0.92 for Bovia. As to the datasets perining to Android, with the Pandex scenario, AVF and ATAE proved the most competitive, yet the highest score was obtained by the latter (0.87). In the Bovia case, OC3 proved the uncontested winner with a score of 0.82 on \verb|ProcessAll|. The closest competitor from AE-APT was the attention-based model ATAE which reached a score of 0.69 on \verb|ProcessEvent|.

Table \ref{tab:winning-algo} summarizes these findings with an emphasis on the highest nDCG scores and the corresponding winner methods. For each OS/attack scenario the max nDCG value is highlighted in red. Observe that AE-APT performed best among the competitors in 7 out of 8 configurations, and demonstrated superior performance, achieving the highest scores in multiple datasets and scenarios.

% =============
\begin{table*}
\centering
%\vspace*{-6 em}%
%\Rotatebox{90}
\small
{%
\begin{tabular}{lllccccc}
\hline
\textbf{OS}&\textbf{Scenario} &\textbf{  }                                                                                                             & 
\textbf{ProcessAll} & \textbf{ProcessEvent} & \textbf{ProcessExec} & \textbf{ProcessParent} & \textbf{ProcessNetflow} \\ \hline
\multirow{4}{*}{\textbf{Cadets (BSD)}}         & \multirow{2}{*}{\textbf{Pandex}} & Highest nDCG   & \cellcolor{red}\textbf{0.87 }      & 0.75         & 0.82        & 0.66          & 0.52             \\
                                               &                                  &Winning method  & \cellcolor{red}ATAE           & LSTMAE            & AE            & ATAE             & VR-ARM              \\ \cline{2-8} 
                                               & \multirow{2}{*}{\textbf{Bovia}}  & Highest nDCG   & \cellcolor{red}\textbf{0.91 }      & 0.54         & 0.85        & 0.80          & 0.60           \\
                                               &                                  & Winning method & \cellcolor{red}AAE          & GRUAE            & AAE           & AAE             & VR-ARM               \\ \hline
\multirow{4}{*}{\textbf{5dir (Windows)}}       & \multirow{2}{*}{\textbf{Pandex}} & Highest nDCG   & 0.79       & \cellcolor{red}\textbf{0.82}         & 0.29        & 0.26            & 0.71             \\
                                               &                                  & Winning method & ATAE          & \cellcolor{red}VR-ARM,ATAE            & AAE           & RNNAE        & OC3              \\ \cline{2-8} 
                                               & \multirow{2}{*}{\textbf{Bovia}}  & Highest nDCG   & \cellcolor{red}0.56       & 0.48         & 0.39        & 0.50         & 0.34           \\
                                               &                                  & Winning method & \cellcolor{red}LSTMAE           & GRUAE             & ATAE            & ATAE              & VR-ARM              \\ \hline
\multirow{4}{*}{\textbf{Trace (Linux)}}        & \multirow{2}{*}{\textbf{Pandex}} & Highest nDCG   & \cellcolor{red}\textbf{0.79}       & 0.74         & 0.46          & 0.24            & 0.54            \\
                                               &                                  & Winning method & \cellcolor{red}ATAE          & ATAE            & ATAE           & OC3             & VR-ARM              \\ \cline{2-8} 
                                               & \multirow{2}{*}{\textbf{Bovia}}  & Highest nDCG   & \cellcolor{red}\textbf{0.92}       & 0.69         & 0.51        & 0.45            & 0.42           \\
                                               &                                  & Winning method & \cellcolor{red}ATAE           & LSTMAE             & ATAE            & ATAE             & ATAE               \\ \hline
\multirow{4}{*}{\textbf{Clearscope (Android)}} & \multirow{2}{*}{\textbf{Pandex}} & Highest nDCG   & 0.83         &\cellcolor{red} \textbf{0.87}           & 0.78        & NA              & 0.67             \\
                                               &                                  & Winning method & AVF,ATAE          & \cellcolor{red}ATAE            & ATAE           & NA              & OC3              \\ \cline{2-8} 
                                               & \multirow{2}{*}{\textbf{Bovia}}  & Highest nDCG   & \cellcolor{red}\textbf{0.82}         & 0.69         & 0.58        & 0.39            & 0.4              \\
                                               &                                  & Winning method & \cellcolor{red}OC3          & ATAE             & LSTMAE            & OC3             & OC3              \\ \hline
\end{tabular}
}
\caption{Highest nDCG scores achieved and the corresponding algorithms on all available datasets. Red highlighted values represent the winner methods with their best nDCG score, for each forensic configuration: OS $\times$ attack scenario $\times$ dataset.}
\label{tab:winning-algo}
\end{table*}

\subsection{Attacks visualization with AE-APT}\label{AE-results}

% %%%%% 

\begin{figure}[t!]
     \centering
    \includegraphics[width=0.5\textwidth]{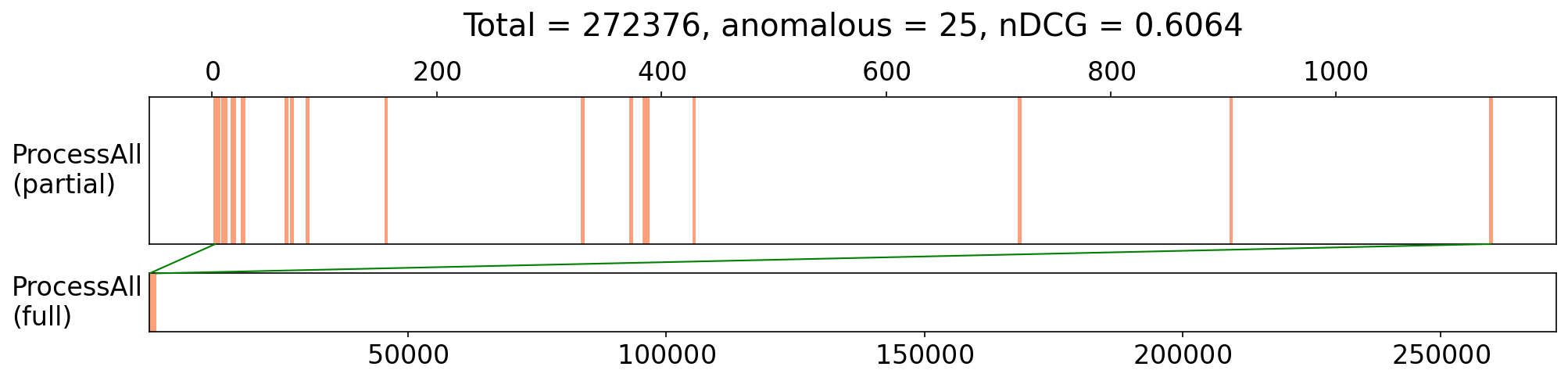}
    \includegraphics[width=0.5\textwidth]{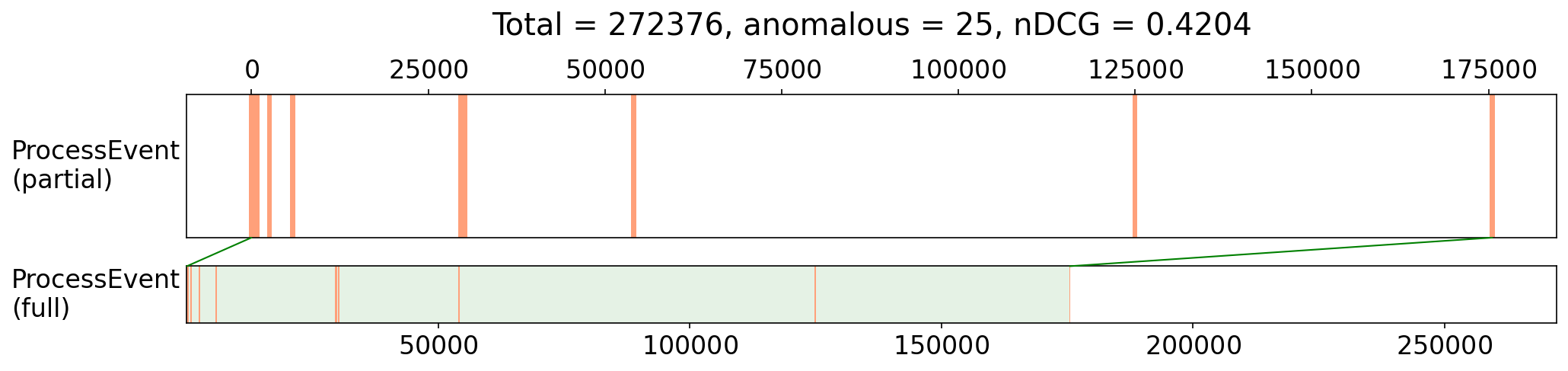}
    \includegraphics[width=0.5\textwidth]{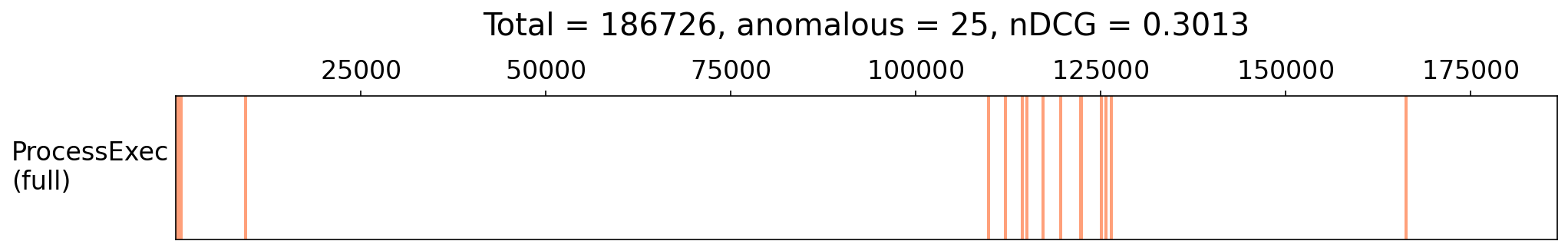}
    \includegraphics[width=0.5\textwidth]{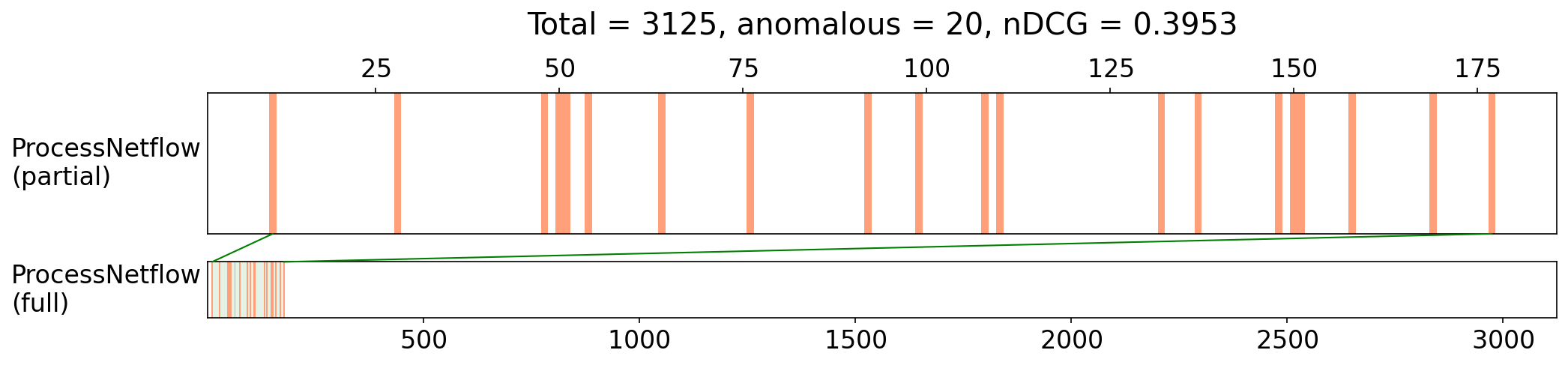}
    \includegraphics[width=0.5\textwidth]{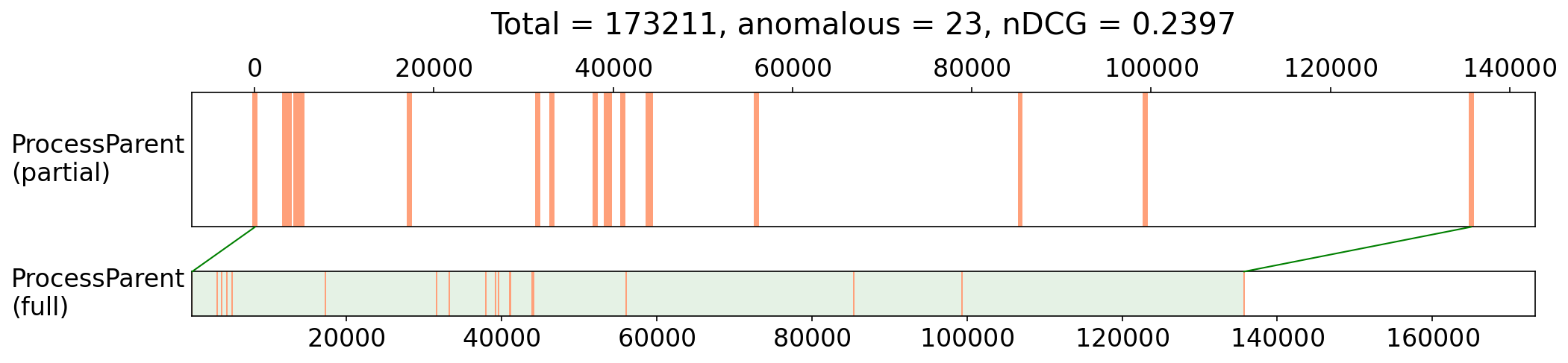}
    \caption{AE AutoEncoder rankings of anomalies on five datasets of the Trace (Linux) system, scenario Pandex.}
    
    \label{Fig:AE_ranking}
\end{figure}

Like a typical AD tool, AE-APT sorts processes in a decreasing order of their anomaly scores. The nDCG score reflects the ranking of the APT-related processes in the resulting sorted list: The higher the ranks, the bigger the score (up till 1). Yet a numerical score hides a substantial part of the ranking information. Hence, we decided to provide some additional visual clues as to how well our models performed. To that end, the following paragraphs delve into the relative performance of two of the above models, the baseline and the adversarial ones (AE and AAE, respectively). For compactness reasons, we restricted the scope of the illustrations to the Pandex scenario on Linux. As a support, we propose a sub-list visualization that covers the interval between the highest and the lowest ranked attack-related process. It is shaped as a horizontal band (see Figure~\ref{Fig:AE_ranking}, for instance) whereby the processes of interest are drawn as orange bars. As such sub-list is typically, but not invariably, much shorter than the total list, a second band is drawn that covers the entire dataset. It works as a zoom-out view of the interval thus providing some contextual information, e.g. where it is located within the full list.

In Figure~\ref{Fig:AE_ranking}, we can observe the ranking visualization produced by the baseline AutoEncoder model for the anomalous processes within the five datasets as well as the model’s nDCG score for that particular dataset. Each subfigure (horizontal band) represents a dataset that is identified on the left-hand side, whereas the $x$-axis (in linear scale) indicates the positions at which the anomalies are ranked by the model. Note that some datasets offer two different views, a global one and a local one, at the top, the latter representing the zoom-in on the interval covering all the attack-related processes. 
Among the five datasets in the figure,
the AE model performs at its best on \verb|ProcessAll|: It ranks all of the anomalies within the top 1200 processes out of the total of 272376. Here, the lowest anomaly rank is 1138, while 4 processes are ranked within the top 10 and 14 within the top 100.

\begin{figure}[t!]
\vspace{-4 em}
    \centering
    \includegraphics[width=0.5\textwidth]{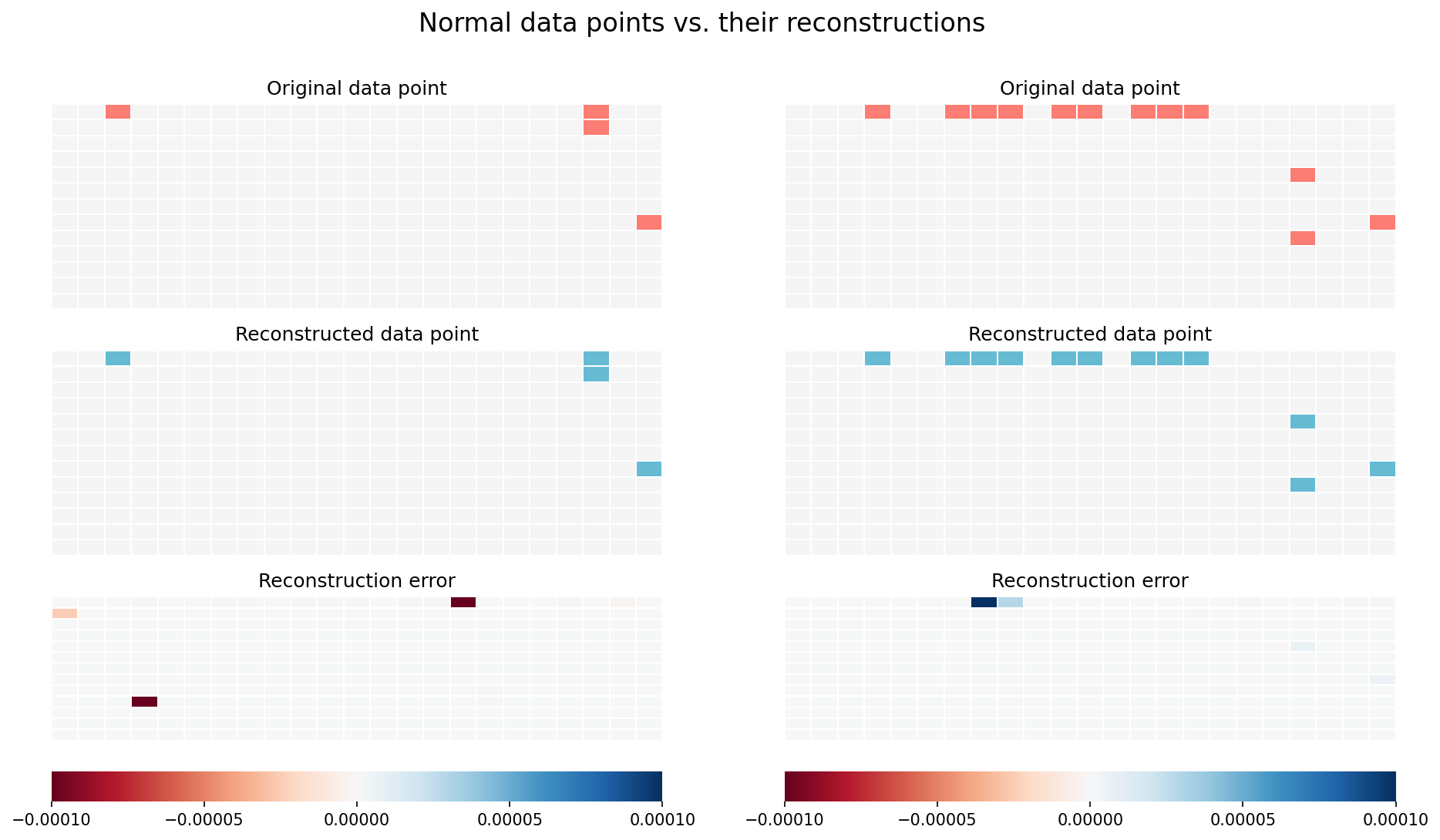}
    \caption{Original data, reconstructed data, and reconstruction error of two random normal data points (AE AutoEncoder).}
    \label{fig:AE_normal_rec}
        \includegraphics[width=0.5\textwidth]{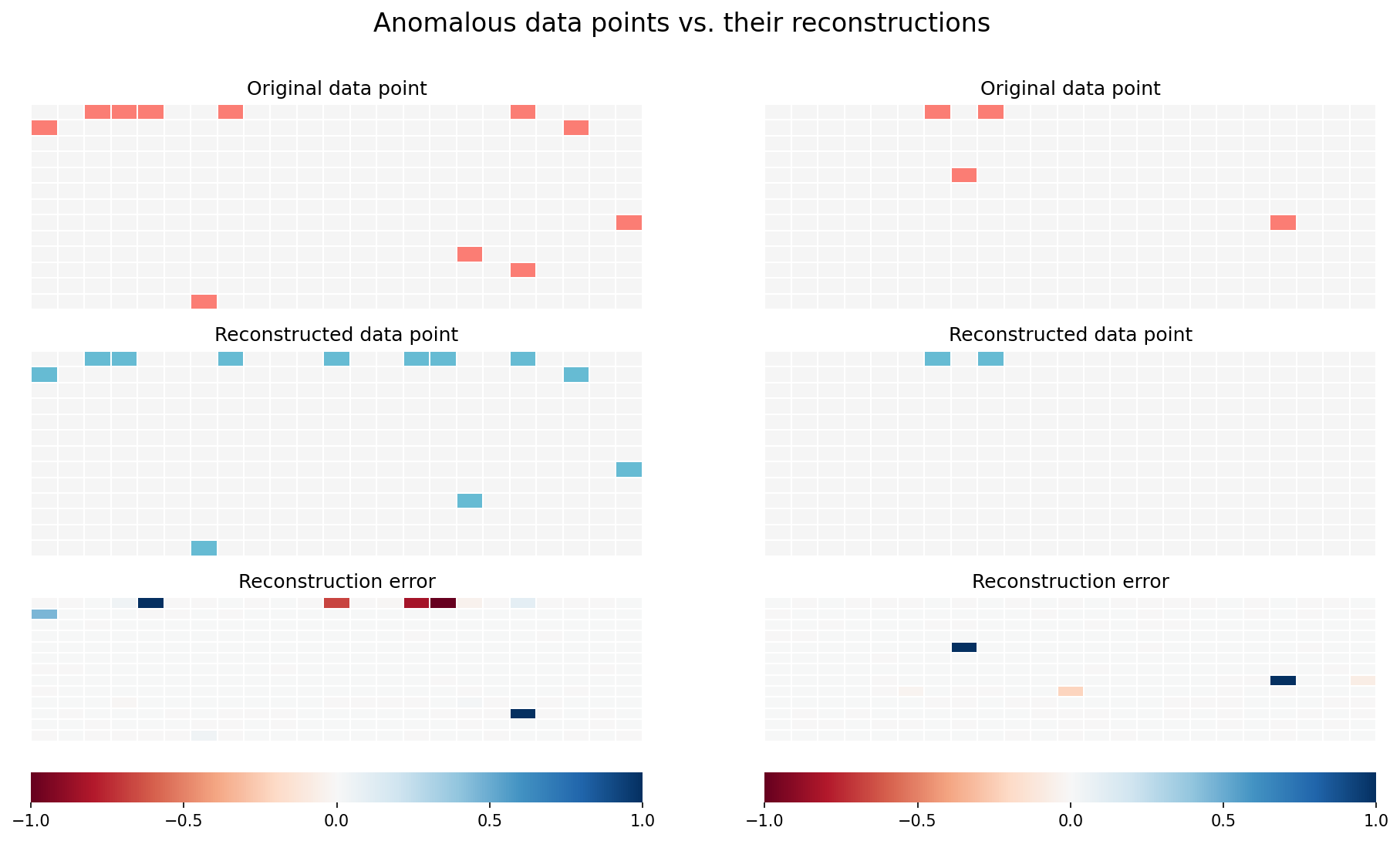}
    \caption{Original data, reconstructed data, and reconstruction error of two random anomalous data points (AE AutoEncoder).}
    \label{fig:AE_anomalous_rec}
\end{figure}

\begin{figure}
\vspace{-6 em}
    \centering
    \includegraphics[width=0.5\textwidth]{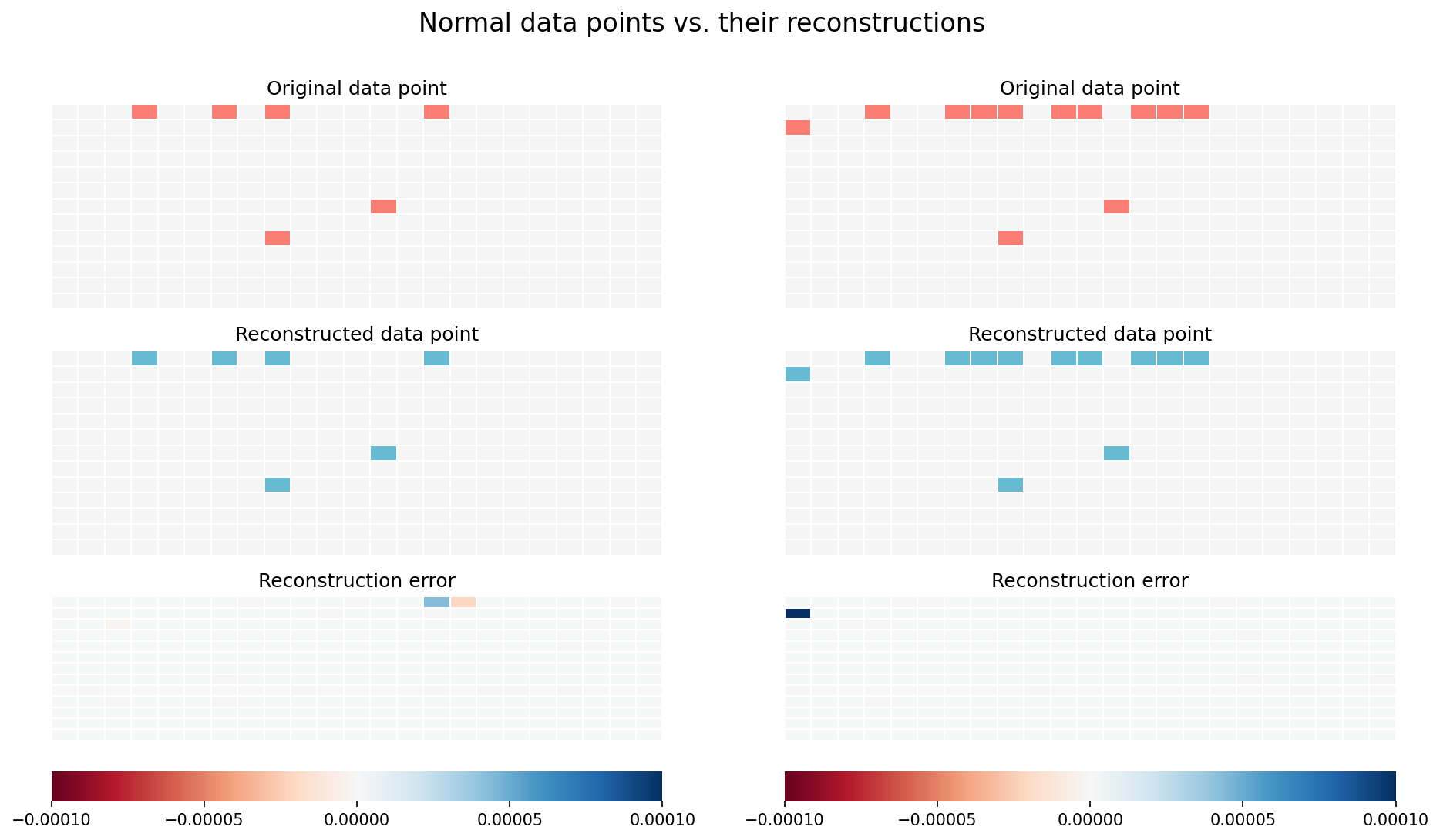}
    \caption{Original data, reconstructed data, and reconstruction error of two random normal data points (AAE Adversarial AutoEncoder).}
    \label{fig:AAE_normal_rec}
        \centering
    \includegraphics[width=0.5\textwidth]{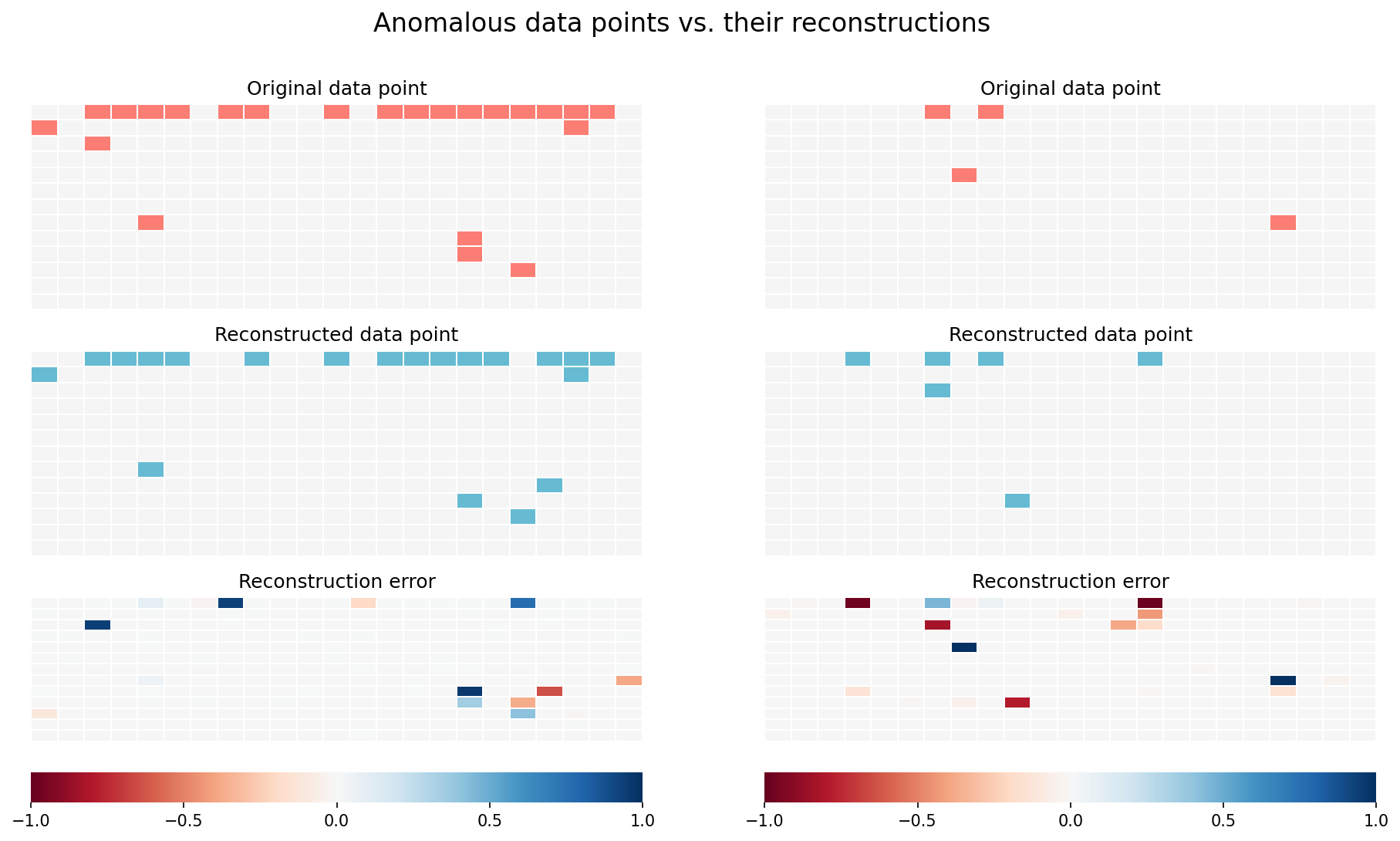}
    \caption{Original data, reconstructed data, and reconstruction error of two random anomalous data points (AAE Adversarial AutoEncoder).}
    \label{fig:AAE_anomalous_rec}
\end{figure}

In a somewhat opposite vein, we illustrate below the very basis for the rankings shown above, that is the support for the anomaly scores. Recall that these boil down to reconstruction errors in the respective AutoEncoder models. Thus, we show in Figure~\ref{fig:AE_normal_rec} the model’s reconstruction of two normal data points sampled from the \verb|ProcessAll| dataset of the Pandex scenario on Linux (recall they are organized into $13 \times 23$ grids). Each point is represented by the 3 diagrams from a column in the figure as follows (same conventions hold for Figures~\ref{fig:AE_anomalous_rec},~\ref{fig:AAE_normal_rec} and~\ref{fig:AAE_anomalous_rec}): The diagrams of the upper tier represent the original data points $x$, those of the middle tier the corresponding reconstruction by the model $\tilde{x}$, and the lower tier shows the error $err = x - \tilde{x}$. 
It is noteworthy that while the original data points are binary vectors ($x[i] \in \{0, 1\}$), their reconstructions are vectors of continuous values ($\tilde{x}[i] \in [0, 1]$), and so are the errors ($err[i] \in [-1, 1]$). Following the same visualization pattern, Figure~\ref{fig:AE_anomalous_rec} depicts the reconstruction of 2 anomalous data points sampled from the same dataset. 
Overall, the model seems to have performed as expected on both data point pairs: In reconstructing normal data points, it does pretty well as it generates small reconstruction errors. Here, they range roughly in $[-0.0001, 0.0001]$. On anomalous data points, our model’s reconstruction is far less accurate, resulting in errors ranging in $[-1, 1]$. We see this as an indication that the model properly learned the essential regularities in normal data and is, therefore, able to properly detect processes showing unusual patterns.

\begin{figure} %[t!]
\vspace{-6 em}
    \centering
    \includegraphics[width=0.5\textwidth]{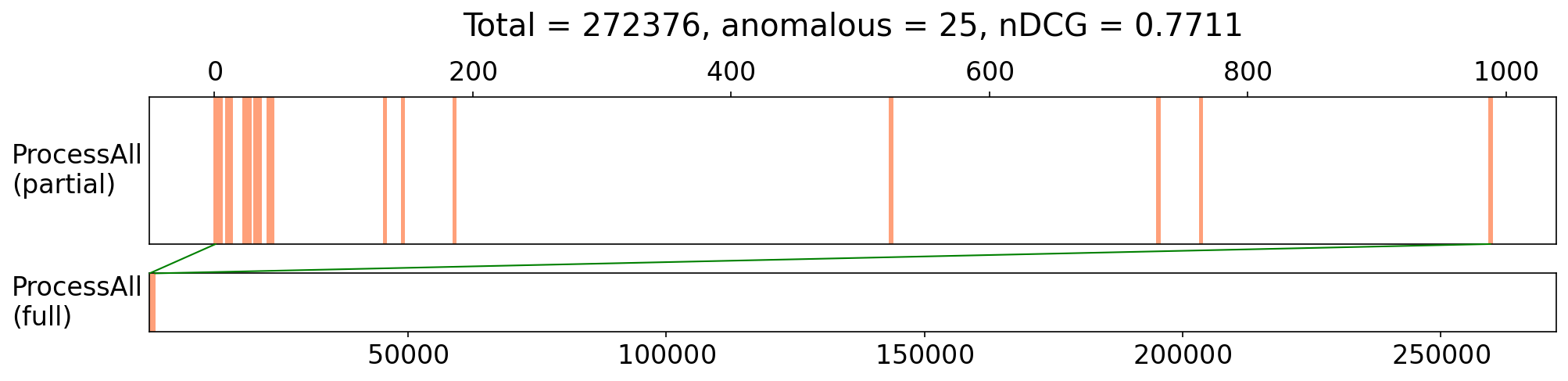}
    \includegraphics[width=0.5\textwidth]{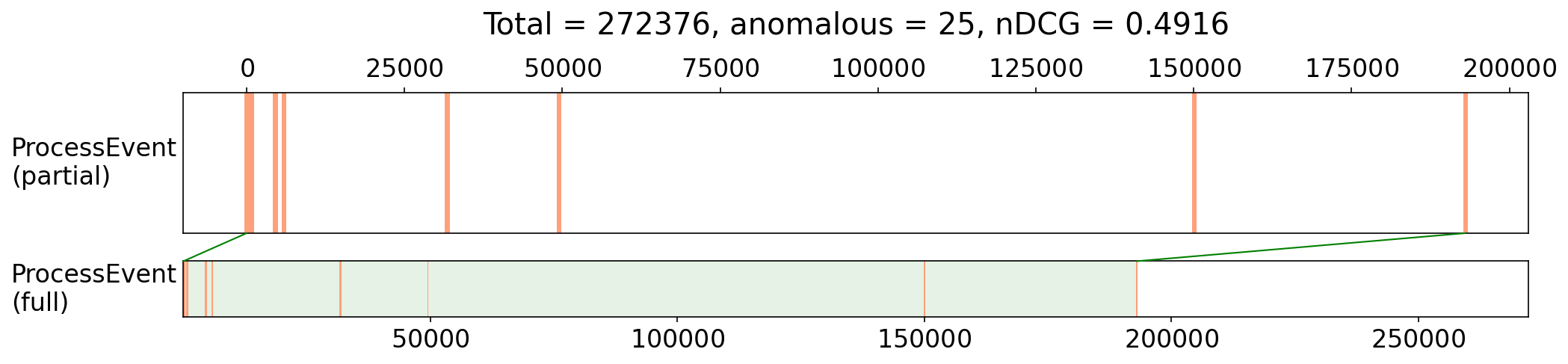}
    \includegraphics[width=0.5\textwidth]{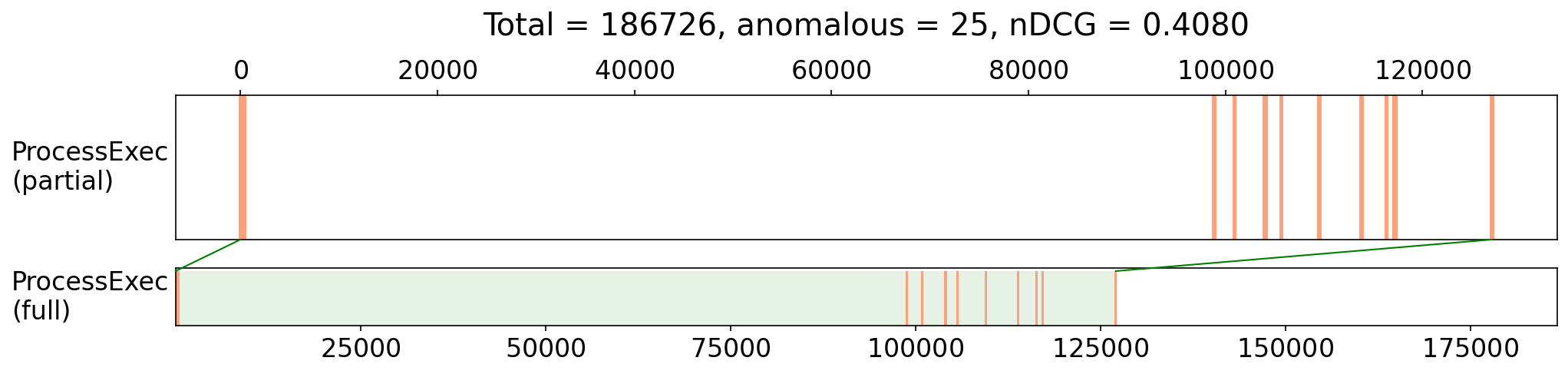}
    \includegraphics[width=0.5\textwidth]{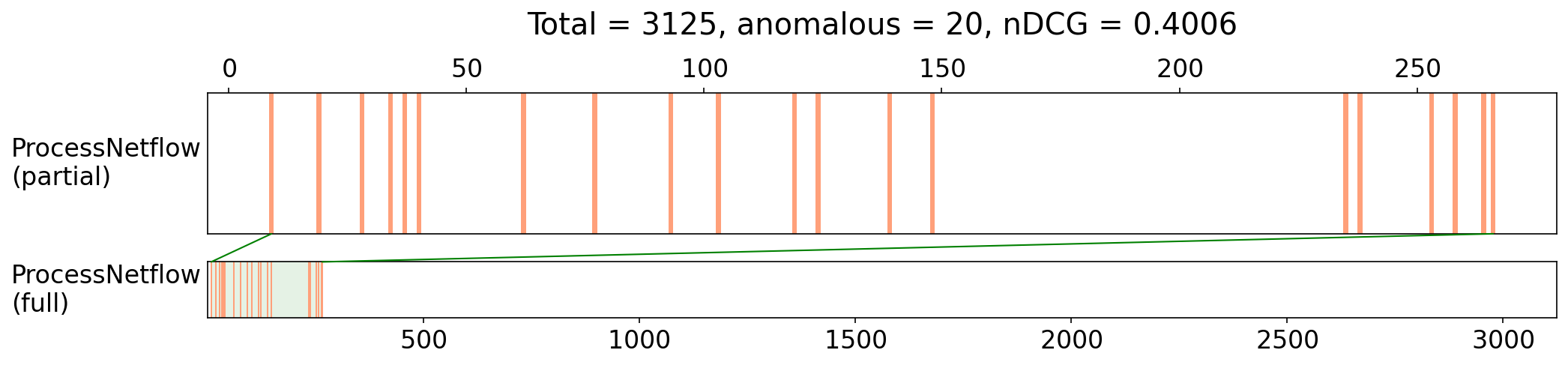}
    \includegraphics[width=0.5\textwidth]{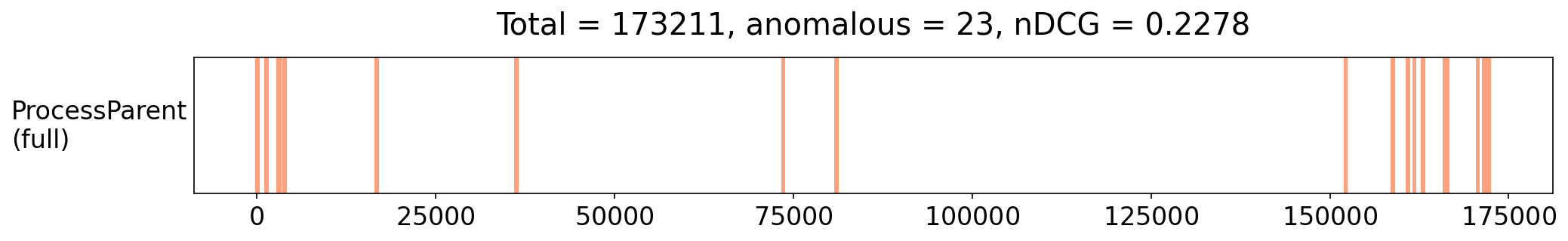}
    \caption{AAE Adversarial AutoEncoder rankings of anomalies on five datasets of the Trace (Linux) system, scenario Pandex.}
    \label{fig:AAE_ranking}
\end{figure}

Next, we focus on the performance of an enhanced AutoEncoder model, the     adversarial one, AAE. As with the baseline AE, we only look at the Pandex scenario on Linux. 
Recall that while the loss functions for the data reconstructing network in AAE, i.e. $\mathbf{G}$, is different from the baseline case, the anomaly score is assigned on the same basis. Again, it's the reconstruction error: Each data point that passes through the model’s generator AutoEncoder and receives the difference $x$ to $\mathbf{G}(x)$ as anomaly score.
Similar to the previous case, Figure~\ref{fig:AAE_ranking} shows the rankings produced by AAE for the anomalous processes in the Panex on Linux dataset suite. The model's top performance is on the \verb|ProcessAll| dataset: All anomalies are ranked under the top 1000 out of a total of 272376 processes  (the lowest-ranked one is at rank 988). Moreover, 5 anomalous processes were ranked among the top 10 and 18 among the top 100.

Next, Figures~\ref{fig:AAE_normal_rec} and~\ref{fig:AAE_anomalous_rec} visualize the model’s reconstruction of data sampled from the top \verb|ProcessAll| dataset (of normal and anomalous  
data points, respectively). The same pattern as in the baseline case can be observed here: While normal data generate few point-wise differences in the diagrams and they remain all within a tight interval centered around 0, anomalies generate much richer drawings with values reaching close to the global limits ($-1$ and $1$, respectively). 

\section{Conclusion}

In this paper, we presented AE-APT, a novel OS-agnostic method for detecting Advanced Persistent Threats (APTs), leveraging a  variety of deep learning models. The proposed framework employs multiple distinct architectures of AutoEncoders to analyze and detect potential anomalies within provenance traces. Specifically, we have implemented six separate types of AutoEncoders, including dense, adversarial, recurrent, and attention-based architectures. 

The dense AutoEncoder focuses on learning a compact representation of the data through fully connected neural networks, which allows it to effectively capture the underlying structure of normal system behavior. The adversarial AutoEncoder, on the other hand, introduces a generative adversarial network (GAN) approach to enhance the robustness of the anomaly detection process by training the model to distinguish between normal and anomalous data more accurately. The recurrent AutoEncoders leverage the strengths of recurrent neural networks (RNNs) to handle the implicit relationship that might exist between system activities. Lastly, the attention-based AutoEncoder incorporates an attention mechanism that enables the model to focus on the most critical parts of the data, thereby improving its ability to identify subtle anomalies that might otherwise go unnoticed.

To validate the effectiveness of AE-APT, we conducted extensive evaluations using a suite of datasets produced as part of the DARPA Transparent Computing program. This includes several large and highly imbalanced datasets, each containing realistic APT scenarios and covering four major operating systems: Windows, Linux, BSD, and Android. The datasets are characterized by their complexity and the inclusion of various forms of APT activities, making them an ideal test stand for assessing the performance of our proposed model.

The results revealed that AE-APT achieves significantly higher detection rates for APTs detection compared to six existing methods. The framework's ability to accurately detect and isolate APTs across different operating systems demonstrates its versatility and robustness. One of the key strengths of AE-APT is its use of the attention mechanism in transformers within the AutoEncoder architecture, which proved particularly effective in enhancing anomaly detection performance. This mechanism allows the model to dynamically weigh the importance of different features, thereby improving its sensitivity to anomalous patterns. In addition to its high detection rates, AE-APT offers advanced visualization mechanisms for anomalies. These visualization tools are designed to foster explainability and interpretability, which are crucial for practical cybersecurity applications. By providing clear and understandable representations of detected anomalies, AE-APT enables cybersecurity professionals to better understand the nature and potential impact of detected threats, facilitating more informed decision-making and response strategies.
Our findings also underscore the potential of attention mechanisms and transformer models in the field of anomaly detection and cybersecurity. The success of AE-APT in detecting APTs across diverse and challenging datasets highlights the value of these advanced deep learning techniques in addressing complex cybersecurity threats. The attention-based approach, in particular, shows promise for further research and development, offering a powerful tool for enhancing the capabilities of anomaly detection systems.

In conclusion, AE-APT represents a significant advancement in the detection of APTs. By leveraging a diverse set of deep learning models and incorporating innovative techniques such as the attention mechanism, our framework provides a robust, flexible, and highly effective solution for identifying and hopefully mitigating APTs across multiple operating systems. The encouraging results obtained from our evaluations demonstrate the potential of AE-APT to improve cybersecurity defenses and protect systems from the evolving threat landscape.

\section{Future Work}

Our future work will focus on refining AE-APT and expanding its capabilities to handle more complex and varied threat scenarios. Key areas of development include model refinement and expansion: We envisage to fine-tune the proposed architectures and explore new deep learning models to enhance detection accuracy and robustness. We would also investigate the benefits of feedback integration along the detection process. Inspired by game theory, we plan to incorporate expert feedback into the model to improve its ranking performance during the classification and increase its capacity to adapt to new and evolving threats. Independently, we will study an online learning-based anomaly detection mode that should further enable real-time analysis and immediate response to ongoing APTs, significantly enhancing the framework’s practical utility. We also aim at enhancing our tool's explainability and interpretability by developing additional visualization tools that foster more effective interface between AI and human experts.

AE-APT, albeit highly effective, might be subject --as any deep learning model-- to some limitations due to dataset specificity and the ensuing overfitting risks. To overcome specificity risks, we aim at expanding and diversifying the training datasets with diffusion models and data augmentation techniques. To prevent overfitting, cross-validation, regularization techniques, and ensemble methods are recommended. Alternatively, domain knowledge about cyber-threats might be infused into the neural models~\cite{sheth2019shades}. 

\section*{Conflicts of Interest}
\vspace{-1 em}
The authors declare no conflict of interest.
\vspace{-1 em}
\section*{Author contributions}
\vspace{-1 em}
SB, NH: Conceptualization, Methodology, Software. SB, JC: Data curation. SB, NH, TR, PV, JC: Visualization, Investigation. SB, NH, PV, TR, JC: Writing.
\vspace{-1 em}
\section*{Acknowledgments}
\vspace{-1 em}
This research was partly funded by the European Research Council (ERC, Skye, 682315). 
\bibliographystyle{ieeetr}\bibliography{sample}

\begin{thebibliography}{10}

\bibitem{sood2012}
A.~K. Sood and R.~J. Enbody, ``Targeted cyberattacks: a superset of advanced
  persistent threats,'' {\em IEEE security and privacy}, vol.~11, no.~1,
  pp.~54--61, 2012.

\bibitem{sujeetha2019cyber}
R.~Sujeetha, H.~Das, T.~Dhelawat, and M.~Tanveer, ``Cyber-space and its
  menaces,'' in {\em 2019 IEEE International Conference on System, Computation,
  Automation and Networking (ICSCAN)}, pp.~1--5, IEEE, 2019.

\bibitem{humayun2020cyber}
M.~Humayun, M.~Niazi, N.~Jhanjhi, M.~Alshayeb, and S.~Mahmood, ``Cyber security
  threats and vulnerabilities: a systematic mapping study,'' {\em Arabian
  Journal for Science and Engineering}, vol.~45, pp.~3171--3189, 2020.

\bibitem{chen2014study}
P.~Chen, L.~Desmet, and C.~Huygens, ``A study on advanced persistent threats,''
  in {\em Communications and Multimedia Security: 15th IFIP TC 6/TC 11
  International Conference, CMS 2014, Aveiro, Portugal, September 25-26, 2014.
  Proceedings 15}, pp.~63--72, Springer, 2014.

\bibitem{alshamrani2019survey}
A.~Alshamrani, S.~Myneni, A.~Chowdhary, and D.~Huang, ``A survey on advanced
  persistent threats: Techniques, solutions, challenges, and research
  opportunities,'' {\em IEEE Communications Surveys \& Tutorials}, vol.~21,
  no.~2, pp.~1851--1877, 2019.

\bibitem{JIA2023110781}
Y.~Jia, Z.~Gu, L.~Du, Y.~Long, Y.~Wang, J.~Li, and Y.~Zhang, ``Artificial
  intelligence enabled cyber security defense for smart cities: A novel attack
  detection framework based on the mdata model,'' {\em Knowledge-Based
  Systems}, vol.~276, p.~110781, 2023.

\bibitem{cole2012advanced}
E.~Cole, {\em Advanced Persistent Threat: Understanding the Danger and How to
  Protect Your Organization}.
\newblock Syngress Publishing, 1st~ed., 2012.

\bibitem{BREWER20145}
R.~Brewer, ``Advanced persistent threats: minimising the damage,'' {\em Network
  Security}, vol.~2014, no.~4, pp.~5--9, 2014.

\bibitem{app13063409}
B.~Genge, P.~Haller, and A.-S. Roman, ``E-aptdetect: Early advanced persistent
  threat detection in critical infrastructures with dynamic attestation,'' {\em
  Applied Sciences}, vol.~13, no.~6, 2023.

\bibitem{SARHAN2021107524}
I.~Sarhan and M.~Spruit, ``Open-cykg: An open cyber threat intelligence
  knowledge graph,'' {\em Knowledge-Based Systems}, vol.~233, p.~107524, 2021.

\bibitem{ghafir2014advanced}
I.~Ghafir, V.~Prenosil, {\em et~al.}, ``Advanced persistent threat attack
  detection: an overview,'' {\em International Journal of Advances in Computer
  Networks and Its Security}, vol.~4, no.~4, p.~5054, 2014.

\bibitem{halbert2016intellectual}
D.~Halbert, ``Intellectual property theft and national security: Agendas and
  assumptions,'' {\em The Information Society}, vol.~32, no.~4, pp.~256--268,
  2016.

\bibitem{shackelford2016protecting}
S.~J. Shackelford, ``Protecting intellectual property and privacy in the
  digital age: the use of national cybersecurity strategies to mitigate cyber
  risk,'' {\em Chapman Law Review}, vol.~19, p.~445, 2016.

\bibitem{ussath2016advanced}
M.~Ussath, D.~Jaeger, F.~Cheng, and C.~Meinel, ``Advanced persistent threats:
  Behind the scenes,'' in {\em 2016 Annual Conference on Information Science
  and Systems (CISS)}, pp.~181--186, IEEE, 2016.

\bibitem{tankard2011advanced}
C.~Tankard, ``Advanced persistent threats and how to monitor and deter them,''
  {\em Network security}, vol.~2011, no.~8, pp.~16--19, 2011.

\bibitem{chen2021few}
M.~Chen, Y.~Wang, H.~Xu, and X.~Zhu, ``Few-shot website fingerprinting
  attack,'' {\em Computer Networks}, vol.~198, p.~108298, 2021.

\bibitem{kshirsagar2023towards}
D.~Kshirsagar and S.~Kumar, ``Towards an intrusion detection system for
  detecting web attacks based on an ensemble of filter feature selection
  techniques,'' {\em Cyber-Physical Systems}, vol.~9, no.~3, pp.~244--259,
  2023.

\bibitem{bhimireddy2023web}
B.~R. Bhimireddy, A.~Nimmagadda, H.~Kurapati, L.~R. Gogula, R.~R. Chintala, and
  V.~C. Jadala, ``Web security and web application security: Attacks and
  prevention,'' in {\em 2023 9th International Conference on Advanced Computing
  and Communication Systems (ICACCS)}, vol.~1, pp.~2095--2096, IEEE, 2023.

\bibitem{FRIEDBERG201535}
I.~Friedberg, F.~Skopik, G.~Settanni, and R.~Fiedler, ``Combating advanced
  persistent threats: From network event correlation to incident detection,''
  {\em Computers \& Security}, vol.~48, pp.~35--57, 2015.

\bibitem{XU2022848}
Y.~Xu, Y.~Fang, C.~Huang, and Z.~Liu, ``Hghan: Hacker group identification
  based on heterogeneous graph attention network,'' {\em Information Sciences},
  vol.~612, pp.~848--863, 2022.

\bibitem{Stuxnet11}
J.~C. Rebane, {\em The Stuxnet Computer Worm and Industrial Control System
  Security}.
\newblock USA: Nova Science Publishers, Inc., 2011.

\bibitem{marczak2018hide}
B.~Marczak, J.~Scott-Railton, S.~McKune, B.~Abdul~Razzak, and R.~Deibert,
  ``Hide and seek: Tracking nso group’s pegasus spyware to operations in 45
  countries,'' tech. rep., 2018.

\bibitem{saad2020attribution}
G.~Saad {\em et~al.}, ``Attribution is in the object: Using rtf object
  dimensions to track apt phishing weaponizers,'' {\em Virus Bulletin},
  vol.~12, pp.~1--2, 2020.

\bibitem{Sakthivelu23}
U.~Sakthivelu and C.~Vinoth~Kumar, ``Advanced persistent threat detection and
  mitigation using machine learning model,'' {\em Intelligent Automation \&
  Soft Computing}, vol.~36, pp.~3691--3707, 01 2023.

\bibitem{denning1987intrusion}
D.~E. Denning, ``An intrusion-detection model,'' {\em IEEE Transactions on
  software engineering}, no.~2, pp.~222--232, 1987.

\bibitem{VIEGAS2017200}
E.~K. Viegas, A.~O. Santin, and L.~S. Oliveira, ``Toward a reliable
  anomaly-based intrusion detection in real-world environments,'' {\em Computer
  Networks}, vol.~127, pp.~200--216, 2017.

\bibitem{aggarwal2017introduction}
C.~C. Aggarwal, {\em An Introduction to Outlier Analysis}.
\newblock Cham: Springer International Publishing, 2017.

\bibitem{chandola2009anomaly}
V.~Chandola, A.~Banerjee, and V.~Kumar, ``Anomaly detection: A survey,'' {\em
  ACM computing surveys (CSUR)}, vol.~41, no.~3, pp.~1--58, 2009.

\bibitem{6890935}
F.~Skopik, G.~Settanni, R.~Fiedler, and I.~Friedberg, ``Semi-synthetic data set
  generation for security software evaluation,'' in {\em 12th Annual
  International Conference on Privacy, Security and Trust}, pp.~156--163, 2014.

\bibitem{MARTINLIRAS2021102202}
L.~F. {Martín Liras} {\em et~al.}, ``Feature analysis for data-driven
  apt-related malware discrimination,'' {\em Computers \& Security}, vol.~104,
  p.~102202, 2021.

\bibitem{lamprakis2017unsupervised}
P.~Lamprakis, R.~Dargenio, D.~Gugelmann, V.~Lenders, M.~Happe, and L.~Vanbever,
  ``Unsupervised detection of apt c\&c channels using web request graphs,'' in
  {\em Detection of Intrusions and Malware, and Vulnerability Assessment: 14th
  International Conference, DIMVA 2017, Bonn, Germany, July 6-7, 2017,
  Proceedings 14}, pp.~366--387, Springer, 2017.

\bibitem{abdullayeva2021advanced}
F.~J. Abdullayeva, ``Advanced persistent threat attack detection method in
  cloud computing based on autoencoder and softmax regression algorithm,'' {\em
  Array}, vol.~10, p.~100067, 2021.

\bibitem{app12136816}
H.~Neuschmied, M.~Winter, B.~Stojanović, K.~Hofer-Schmitz, J.~Božić, and
  U.~Kleb, ``Apt-attack detection based on multi-stage autoencoders,'' {\em
  Applied Sciences}, vol.~12, no.~13, 2022.

\bibitem{9496635}
B.~Min, J.~Yoo, S.~Kim, D.~Shin, and D.~Shin, ``Network anomaly detection using
  memory-augmented deep autoencoder,'' {\em IEEE Access}, vol.~9,
  pp.~104695--104706, 2021.

\bibitem{AHMED201619net}
M.~Ahmed {\em et~al.}, ``A survey of network anomaly detection techniques,''
  {\em Journal of Network and Computer Applications}, vol.~60, pp.~19--31,
  2016.

\bibitem{mchugh2000testing}
J.~McHugh, ``Testing intrusion detection systems: a critique of the 1998 and
  1999 darpa intrusion detection system evaluations as performed by lincoln
  laboratory,'' {\em ACM Transactions on Information and System Security
  (TISSEC)}, vol.~3, no.~4, pp.~262--294, 2000.

\bibitem{mahoney2003analysis}
M.~V. Mahoney and P.~K. Chan, ``An analysis of the 1999 darpa/lincoln
  laboratory evaluation data for network anomaly detection,'' in {\em
  International Workshop on Recent Advances in Intrusion Detection},
  pp.~220--237, Springer, 2003.

\bibitem{ALJAWARNEH2018152}
S.~Aljawarneh {\em et~al.}, ``Anomaly-based intrusion detection system through
  feature selection analysis and building hybrid efficient model,'' {\em
  Journal of Computational Science}, vol.~25, pp.~152--160, 2018.

\bibitem{shafi2013evaluation}
K.~Shafi and H.~A. Abbass, ``Evaluation of an adaptive genetic-based signature
  extraction system for network intrusion detection,'' {\em Pattern Analysis
  and Applications}, vol.~16, no.~4, pp.~549--566, 2013.

\bibitem{auty2015anatomy}
M.~Auty, ``Anatomy of an advanced persistent threat,'' {\em Network Security},
  vol.~15, no.~4, pp.~13--16, 2015.

\bibitem{streamspot}
E.~Manzoor, S.~Milajerdi, {\em et~al.}, ``Fast memory-efficient anomaly
  detection in streaming heterogeneous graphs,'' in {\em Proceedings of the
  22nd ACM SIGKDD international conference on knowledge discovery and data
  mining}, pp.~1035--1044, 2016.

\bibitem{han2018tapp}
X.~Han, T.~Pasquier, and M.~Seltzer, ``Provenance-based intrusion detection:
  Opportunities and challenges,'' in {\em 10th USENIX Workshop on the Theory
  and Practice of Provenance (TaPP 2018)}, 2018.

\bibitem{unicorn}
X.~Han {\em et~al.}, ``Unicorn: Runtime provenance-based detector for advanced
  persistent threats,'' in {\em {NDSS}}, 2020.

\bibitem{BerradaCBMMTW20}
G.~Berrada, J.~Cheney, S.~Benabderrahmane, {\em et~al.}, ``A baseline for
  unsupervised advanced persistent threat detection in system-level
  provenance,'' {\em Future Generation Computer Systems}, vol.~108,
  pp.~401--413, 2020.

\bibitem{Benabderrahmane21}
S.~Benabderrahmane, G.~Berrada, J.~Cheney, and P.~Valtchev, ``A rule
  mining-based advanced persistent threats detection system,'' in {\em Proc of
  {IJCAI}21, Montreal, Canada} (Z.~Zhou, ed.), pp.~3589--3596, ijcai.org, 2021.

\bibitem{awad2016data}
A.~Abir, S.~Kadry, {\em et~al.}, ``Data leakage detection using system call
  provenance,'' in {\em 2016 International Conference on Intelligent Networking
  and Collaborative Systems (INCoS)}, pp.~486--491, 2016.

\bibitem{jenkinson2017applying}
Jenkinson {\em et~al.}, ``Applying provenance in {APT} monitoring and analysis:
  Practical challenges for scalable, efficient and trustworthy distributed
  provenance,'' in {\em TaPP}, pp.~16--21, 2017.

\bibitem{koufakou_2007}
A.~Koufakou, E.~G. Ortiz, {\em et~al.}, ``A scalable and efficient outlier
  detection strategy for categorical data,'' in {\em 19th {IEEE} Int Conf on
  Tools with Artificial Intelligence({ICTAI} 2007)}, pp.~210--217, {IEEE}, Oct.
  2007.

\bibitem{he_2005}
Z.~He {\em et~al.}, ``Fp-outlier: Frequent pattern based outlier detection,''
  {\em Comput. Sci. Inf. Syst.}, vol.~2, pp.~103--118, 2005.

\bibitem{narita_2008}
K.~Narita and H.~Kitagawa, ``Outlier detection for transaction databases using
  association rules,'' in {\em 2008 The 9th Int Conf on Web-Age Information
  Management}, pp.~373--380, 2008.

\bibitem{smets2011}
K.~Smets and J.~Vreeken, ``The odd one out: Identifying and characterising
  anomalies,'' in {\em Proceedings of the 2011 SIAM international conference on
  data mining}, pp.~804--815, SIAM, 2011.

\bibitem{pang_2020}
G.~Pang {\em et~al.}, ``Deep learning for anomaly detection: {A} review,'' {\em
  CoRR}, vol.~abs/2007.02500, 2020.

\bibitem{goodfellow2014}
I.~e.~a. Goodfellow, ``Generative adversarial nets,'' in {\em Advances in
  Neural Information Processing Systems} (Z.~Ghahramani, M.~Welling, C.~Cortes,
  N.~Lawrence, and K.~Weinberger, eds.), vol.~27, Curran Associates, Inc.,
  2014.

\bibitem{pascanu2013construct}
R.~Pascanu, C.~Gulcehre, K.~Cho, and Y.~Bengio, ``How to construct deep
  recurrent neural networks,'' {\em arXiv preprint arXiv:1312.6026}, 2013.

\bibitem{hochreiter1997long}
S.~Hochreiter and J.~Schmidhuber, ``Long short-term memory,'' {\em Neural
  computation}, vol.~9, no.~8, pp.~1735--1780, 1997.

\bibitem{chung2014empirical}
J.~Chung, C.~Gulcehre, K.~Cho, and Y.~Bengio, ``Empirical evaluation of gated
  recurrent neural networks on sequence modeling,'' {\em arXiv preprint
  arXiv:1412.3555}, 2014.

\bibitem{bahdanau2016neural}
D.~Bahdanau, K.~Cho, and Y.~Bengio, ``Neural machine translation by jointly
  learning to align and translate,'' 2016,arXiv:1409.047.

\bibitem{vaswani2023attention}
A.~Vaswani, N.~Shazeer, N.~Parmar, J.~Uszkoreit, L.~Jones, A.~N. Gomez,
  {\L}.~Kaiser, and I.~Polosukhin, ``Attention is all you need,'' {\em Advances
  in neural information processing systems}, vol.~30, 2017.

\bibitem{darpa}
``Transparent computing.'' https://www.darpa.mil/program/transparent-computing.

\bibitem{berrada_2019}
G.~Berrada and J.~Cheney, ``Aggregating unsupervised provenance anomaly
  detectors,'' in {\em 11th International Workshop on Theory and Practice of
  Provenance (TaPP 2019)}, (Philadelphia, PA), USENIX Association, June 2019.

\bibitem{jarvelin_2002}
K.~J\"{a}rvelin and J.~Kek\"{a}l\"{a}inen, ``Cumulated gain-based evaluation of
  {IR} techniques,'' {\em {ACM} Transactions on IS}, vol.~20, pp.~422--446,
  Oct. 2002.

\bibitem{sheth2019shades}
A.~Sheth, M.~Gaur, U.~Kursuncu, and R.~Wickramarachchi, ``Shades of
  knowledge-infused learning for enhancing deep learning,'' {\em IEEE Internet
  Computing}, vol.~23, no.~6, pp.~54--63, 2019.

\end{thebibliography}

\end{document}